\documentclass{svjour3}                     % onecolumn (standard format)
\smartqed  % flush right qed marks, e.g. at end of proof
\usepackage{graphicx}

\usepackage{epsfig}
\usepackage{amssymb}
\usepackage{epstopdf}
\newcommand \Pomeron {I\!\!P}

\begin{document} 

\title{Disentangling coherent and incoherent quasielastic $J/\psi$ photoproduction on nuclei by neutron tagging 
in ultraperipheral ion collisions at the LHC}

\author{V.~Guzey \and M.~Strikman \and M.~Zhalov}

\institute{V.~Guzey \at Petersburg Nuclear Physics Institute (PNPI), National Research 
Center ``Kurchatov Institute'', Gatchina, 188300, Russia\\
\email{vguzey@jlab.org}
\and M.~Strikman \at
The Pennsylvania State University, University Park, PA 16802, USA\\
\email{strikman@phys.psu.edu}
\and M.~Zhalov \at
Petersburg Nuclear Physics Institute (PNPI), National Research 
Center ``Kurchatov Institute'', Gatchina, 188300, Russia\\
\email{zhalov@pnpi.spb.ru}
}

\date{Received: date / Accepted: date}

\maketitle

\begin{abstract}

We consider $J/\psi$ photoproduction in ion--ion ultraperipheral collisions (UPCs) at the LHC and RHIC 
in the coherent and incoherent quasielastic
channels with and without accompanying forward neutron emission and analyze the role
of nuclear gluon shadowing at small $x$, 
$x=10^{-4}-10^{-2}$, 
in these processes.
We find that despite the good agreement between large nuclear gluon shadowing and
the ALICE data in the coherent channel, in the incoherent channel,
the leading twist approximation predicts the amount of nuclear suppression 
which is by approximately a factor of $1.5$ exceeds that seen in the data.
We hypothesize that part of the discrepancy
can be  accounted for by the incoherent inelastic process of $J/\psi$ photoproduction with nucleon dissociation.
To separate the high-photon-energy and low-photon-energy contributions to the 
$d \sigma_{AA\to AAJ/\psi}(y)/dy$ cross section, we consider ion--ion UPCs accompanied by
neutron emission due to electromagnetic excitation of one or both colliding nuclei. 
We describe the corresponding PHENIX data and make predictions for the LHC kinematics.
In addition, in the incoherent quasielastic case, 
we show that the separation between the low-photon-energy and
high-photon-energy contributions can be efficiently performed by measuring the correlation between 
the directions of $J/\psi$ and the emitted neutrons.


\end{abstract}

\section{Introduction}
\label{sec:Introduction}

Recently coherent and incoherent photoproduction of $J/\psi$ in ultraperipheral collisions (UPCs) of nuclei 
was  measured by the ALICE collaboration at the LHC~\cite{Abbas:2013oua,Abelev:2012ba}.
In the coherent channel,
a large reduction of the coherent cross section---approximately by a factor of three---as compared  
to the impulse approximation
has been reported. 
Such a magnitude of the suppression was found to be in the 
reasonable agreement with the expectations 
of the approaches predicting significant nuclear gluon shadowing at 
%vg2014 $x =10^{-3}-10^{-2}$
 $x \approx 10^{-3}$
($x$ is the  fraction of the nucleus momentum carried by gluons),
notably with predictions 
of the leading twist approach to nuclear shadowing~\cite{Guzey:2013xba,Guzey:2013qza} and
with the results of the EPS09 global QCD fit to nuclear parton distributions~\cite{Eskola:2009uj}.
Thus, charmonium photoproduction on nuclear targets is a useful tool to study 
nuclear gluon shadowing at small $x$.

The aim of this paper is twofold. First, we extend application of the leading twist approach to nuclear 
shadowing~\cite{Frankfurt:2011cs} to incoherent quasielastic  photoproduction of $J/\psi$ on nuclei 
and show that the suppression  of both coherent and
incoherent $J/\psi$ photoproduction in ion--ion UPCs 
can be described in the same framework. A comparison of the resulting theoretical 
prediction for the cross section of incoherent  $J/\psi$ photoproduction in Pb-Pb UPCs at the LHC 
to the ALICE data, 
%vg2014
which is also characterized to correspond to an incoherent quasielastic process~\cite{Abbas:2013oua},
 shows that the expected suppression due to nuclear shadowing 
is larger than that seen in the data. 
We argue that this does not only place additional constraints on models of nuclear 
shadowing down to 
$x \approx 10^{-4}$
but also indicates that 
%vg
additional processes can contribute to the ALICE data.
%contributions (such as, e.g., the process of 
%nucleon dissociation) play an important role in the incoherent channel.
%vg2014.
% -------------------------------------------------
In particular, on top of incoherent $J/\psi$ photoproduction on nuclei resulting from the target 
nucleus excitation, the $\gamma+A \to J/\psi+Y+(A-1)^{\ast}$ process driven by the $\gamma+N \to J/\psi+Y$ 
nucleon dissociation
($Y$ denotes products of the nucleon dissociation) accompanied by the nucleus breakup into the $(A-1)^{\ast}$ 
system consisting of nucleus debris or nucleons
also leads to the inelastic final state. 
The calculation of the $\gamma+A \to J/\psi+Y+(A-1)^{\ast}$ contribution is rather involved
reflecting different mechanisms of the elementary reaction at 
small and large $|t|$ considered in~\cite{Frankfurt:2008et}
 and 
will be addressed in a separate publication.
% ----

Second, we discuss specifics of and make predictions for
coherent and incoherent charmonium production in nucleus--nucleus UPCs
accompanied by forward neutron emission which can be studied
at the LHC with the ALICE, CMS and ATLAS detectors equipped by zero degree calorimeters (ZDC).  
The following channels can be studied:
(i) one of the nuclei emits at least one neutron 
while its partner does not --- (0nXn); (ii) both nuclei emit neutrons in opposite
directions --- (XnXn), (iii) neither of the nuclei emits neutrons  --- (0n0n). 
We show that selection of a specific channel can strongly influence
the ratio of the cross sections of
incoherent to coherent $J/\psi$ photoproduction in Pb-Pb UPCs at the LHC.  
In particular, 
we argue that
the study of incoherent production of 
charmonium in ion--ion UPCs with the nucleus breakup allows one to separate 
the low-photon-energy and high-photon-energy contributions to nuclear 
$J/\psi$ photoproduction 
and, hence, to provide additional 
information on the dynamics of 
nuclear shadowing of the gluon distribution in nuclei which is  
complementary to that obtained from coherent onium production.

This paper is organized as follows. In Sect.~\ref{sec:shadowing}, we  
discuss the suppression of the coherent and incoherent nuclear $J/\psi$ 
photoproduction cross sections due to nuclear shadowing. 
We briefly recapitulate main results of the vector meson dominance and the color dipole models for these
processes and present the derivation of the coherent $\sigma_{\gamma A \to J/\psi A}$ and the
incoherent $\sigma_{\gamma A \to J/\psi A^{\prime}}$ cross sections in the leading twist approximation. 
In Sect.~\ref{sec:UPCs}, using the results of Sect.~\ref{sec:shadowing}, we make predictions for 
the coherent and incoherent cross section of $J/\psi$ photoproduction in ion--ion UPCs without and with neutron
emission and analyze the obtained results.
 Section~\ref{sec:conclusions} presents a brief summary of the obtained results.

\section{Nuclear gluon shadowing in coherent and incoherent $J/\psi$ 
photoproduction on nuclei}
\label{sec:shadowing}

\subsection{The coherent nuclear $J/\psi$ photoproduction cross section}
\label{subsec:coherent_case}

The earliest model for production of vector mesons off nuclei 
is the vector meson dominance model based on hadronic degrees of freedom~\cite{Sakurai_1969}.
 In the high-energy optical limit of the Glauber model,
the standard expression for the cross section of coherent $J/\psi$ photoproduction
on a nuclear target reads~(see, e.g.,~\cite{Bauer:1977iq}):
\begin{equation}
\sigma_{\gamma A\rightarrow J/\psi A}^{\rm VMD}(W_{\gamma p})=
\frac{d\sigma_{\gamma p\rightarrow J/\psi p}(W_{\gamma p},t=0)}{dt}
\left[{\frac {\sigma_{J/\psi A}^{\rm tot} (W_{\gamma p}) } 
 {A\sigma_{J/\psi N}^{\rm tot} (W_{\gamma p}) }}\right]^2  \Phi_A(t_{\rm min})
 \,,
\label{sigglc}
\end{equation}
where we assumed that the multiple interactions 
leading to the nuclear shadowing effect 
do not distort the shape of the transverse momentum
distribution of the vector meson.
In Eq.~(\ref{sigglc}),  
$\Phi_A(t_{\rm min})=\int \limits^{t_{\rm min}}_{-\infty} dt \left|F_A (t)\right|^2$, 
where $F_A (t)$ is the nucleus form factor 
(its normalization is $F_A (0)=A$) 
whose square takes into account
that the nucleus remains intact in the coherent process, 
$t$ is the four-momentum transfer squared 
 and 
$t_{\rm min}=-M_{J/\psi}^4 m_N^2/W_{\gamma p}^4$ is 
its minimal value
($M_{J/\psi}$ and $m_N$ are the masses of $J/\psi$ and the nucleon, respectively);
$W_{\gamma p}$ is the photon--nucleus center of mass energy per nucleon;
$\sigma_{J/\psi A}^{\rm tot}$ and $\sigma_{J/\psi N}^{\rm tot}$ are the total $J/\psi$--nucleus and  
$J/\psi$--nucleon cross sections, respectively. Note that in 
the first and second terms in Eq.~(\ref{sigglc}), the dependence on $t_{\rm min}$ has been safely 
neglected compared to the $\Phi_A(t_{\rm min})$ term.

In the optical limit of the Glauber model, the $\sigma_{J/\psi A}^{\rm tot}$  cross section is:
 \begin{equation}
\sigma_{J/\psi A}^{\rm tot}(W_{\gamma p})=2 \int d^2{\vec b}
\left[1-\exp\left\{-{{\sigma_{J/\psi N}(W_{\gamma p})T_A ({\vec b})}\over 2}
\right\}\right] \,,
\label{gltsig}
\end{equation}
where $T_A (\vec b) =\int d^2 \vec b \rho_A (\vec b,z)$
is the nuclear width function; 
$\rho_A(\vec b,z)$ is the nuclear density.
Equation~(\ref{gltsig}) describes successive multiple interactions of $J/\psi$ with target nucleons, whose
destructive interference results in the nuclear attenuation (shadowing)~\cite{Glauber:1955qq} 
of the $J/\psi$--nucleus 
cross section, $\sigma_{J/\psi A}^{\rm tot} < A \sigma_{J/\psi N}^{\rm tot}$.

The main issue with Eqs.~(\ref{sigglc}) and (\ref{gltsig})
is what value of the 
elementary 
$\sigma_{J/\psi N}^{\rm tot}$ cross section to use.
It has been well known for a long time that if one tries to determine
$\sigma_{J/\psi N}^{\rm tot}$ using the vector meson dominance model
and  the data on the elementary
$\gamma p\rightarrow J/\psi p$ process, the obtained value of
$\sigma_{J/\psi N}^{\rm tot}$ is small, namely, 
$\sigma_{J/\psi N}^{\rm tot}(W_{\gamma p}=5\, {\rm GeV})  \approx 1$ mb 
and $\sigma_{J/\psi N}^{\rm tot}(W_{\gamma p}=100\, {\rm GeV}) \approx 3$ mb, see, e.g., \cite{Hufner:1997jg}.
As a result, the effect of nuclear shadowing in the 
$\sigma_{\gamma A\rightarrow J/\psi A}^{\rm VMD}$ cross section
predicted using Eqs.~(\ref{sigglc}) and (\ref{gltsig})
turns out to be small
%vg2014
in the small-$x$ region,
which contradicts the ALICE data.
Also, the smallness of $\sigma_{J/\psi N}^{\rm tot}$ 
serves as an indication that in the strong interaction, $J/\psi$ reveals 
properties of a small-size dipole built from a heavy quark-antiquark pair.

The simple space--time picture of heavy onium production in the vector meson dominance model
is superseded by the one in high-energy QCD, where 
the process of charmonium photoproduction involves 
three stages: (i) the photon conversion into a $q\bar q$ 
component (dipole) long before the target, (ii) the interaction of the dipole
with the target, and (iii) the conversion of
the $q\bar q$ component into the final state vector meson.
This space--time picture is properly taken into 
account/realized
in the framework of 
the QCD dipole approximation~\cite{Nikolaev:1990ja,Mueller:1994jq}, where
the large mass of the $c$-quark, $m_c$, sets the hard scale in 
diffractive charmonium photoproduction, $\mu^2 \sim {\cal O}(m_c^2)$, and thus,
leads to the dominance of the dipoles of the small transverse size $d_t$,  $d_t^2 \sim 1/\mu^2$. 
The corresponding $q{\bar q}$ dipole--target cross section 
reads~\cite{Blaettel:1993rd}:
\begin{equation}
\sigma_{q{\bar q}T}(d_t^2,x)=\frac{\pi^2}{3}d_t^2
\alpha_s({\bar Q}^2)xG_T(x,{\bar Q}^2) \,,
\label{eq:sigma_dipole}
\end{equation}
where $\alpha_s$ is the running strong coupling constant; $G_T(x,{\bar Q}^2)$ is the gluon 
distribution in the target $T$, which depends on the momentum fraction 
 $x=M_{J/\psi}^2/W_{\gamma p}^2$ ($M_{J/\psi}$ is the mass of $J/\psi$)
and the hard scale squared
 ${\bar Q}^2 \propto 1/d_t^2 \sim {\cal O}(m_c^2)$.

In general, in the dipole approach, 
the $J/\psi$ photoproduction cross section involves dipoles of different transverse
sizes $d_t$, which corresponds to different scales ${\bar Q}^2$ in Eq.~(\ref{eq:sigma_dipole}).
However,  using the leading logarithmic approximation
and making 
simplifying assumptions about the gluon transverse momentum ${\tilde k}_t$ (${\tilde k}_t^2 \ll \mu^2$)
and the $J/\psi$ wave function (assuming that the transverse momenta of $c$-quarks, $k_t$, are small 
($k_t^2 \ll m_c^2$),
while the longitudinal ones are equally shared between the $c$-quark and antiquark),
it  was shown that the imaginary part of the 
$J/\psi$ photoproduction amplitude
involving the dipole cross section of Eq.~(\ref{eq:sigma_dipole})  
is expressed through the gluon density of the target at the scale of
$\mu^2=M_{J/\psi}^2/4=2.4$ GeV$^2$ and one obtains~\cite{Ryskin:1992ui}:
\begin{equation}
\frac {d\sigma_{\gamma T\rightarrow J/\psi T }(W_{\gamma p},t=0)} {dt}=C({\mu}^2)
\left[xG_T (x,\mu^2)\right]^2 \,,
\label{eq:sigma_ryskin}
\end{equation}
where $C(\mu^2)=(1+\eta^2)R_g^2 \,F^2(\mu^2)M^3_{J/\psi}\Gamma_{ee} {\pi}^3 {\alpha_{s}}^2(\mu^2 
)/(48{\alpha_{\rm e.m.}}\mu^8)$;
$\Gamma_{ee}$ is the width of the $J/\psi$ electronic decay
and $\alpha_{\rm e.m.}$ is the fine-structure constant. 
The factors of $\eta$ and $R_g$ correct  Eq.~(\ref{eq:sigma_ryskin}) for the real part and 
the skewness of the $\gamma T \to J/\psi T$ scattering amplitude, respectively; 
the factor of $F^2(\mu^2)$ absorbs all effects not included in the used approximation
($F^2(\mu^2)=1$ in the non-relativistic limit for the $J/\psi$ wave function).

In a more general case~\cite{Brodsky:1994kf}, (i.e., beyond the 
$k_t^2 \ll m_c^2$ limit
for the $J/\psi$ wave function),
there exists a theoretical uncertainty in the value of $\mu^2$ in Eq.~(\ref{eq:sigma_ryskin})
which means that one could use a reasonable range of values, e.g., $\mu^2=2.4-3.4$ GeV$^2$. 
For example, the suitable value of $\mu^2$ can be determined phenomenologically~\cite{Guzey:2013qza} 
comparing predictions of Eq.~(\ref{eq:sigma_ryskin}) for the proton target
with the data.

\begin{figure}[htbp]
\centering
\epsfig{file=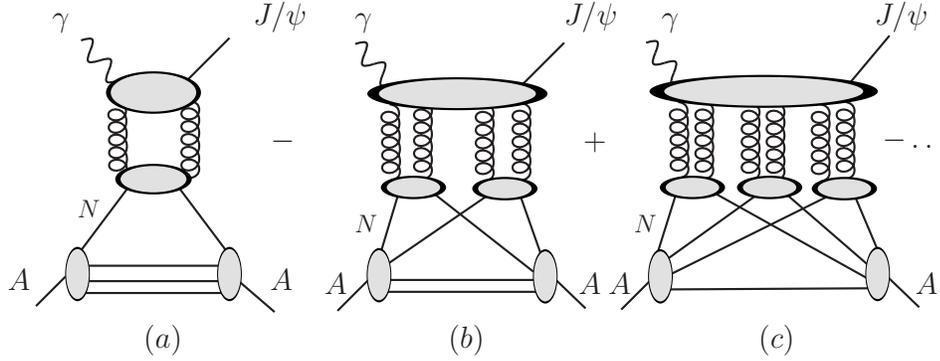,scale=0.8}
\caption{The multiple scattering series for the $\gamma A \to J/\psi A$ scattering amplitude in the color dipole formalism: (a) the impulse approximation,  (b) the double scattering, (c) the interaction with three nucleons of the target.
}
\label{fig:Series_Jpsi_eikonal}
\end{figure}

Working in the framework of the color dipole model, one can calculate the  
$\gamma A \to J/\psi A$ scattering amplitude by summing multiple rescatterings of a dipole of 
the fixed size $d_t$ on the target nucleons essentially using Eq.~(\ref{gltsig}) and then integrating over
$d_t$ with the weight given by the photon and $J/\psi$ wave 
functions~\cite{Kopeliovich:2001xj}, see Fig.~\ref{fig:Series_Jpsi_eikonal}.
Since, on average, $J/\psi$ photoproduction is dominated by small transverse size dipoles
and the corresponding dipole--nucleon cross section~(\ref{eq:sigma_dipole}) is small, the resulting
nuclear shadowing is also small~\cite{Goncalves:2011vf,Lappi:2013am}, which contradicts the ALICE data, 
see the discussion in~\cite{Guzey:2013qza}.
In this respect, the situation is similar to the VMD case considered above.
In other words, in the dipole formalism,
when only $q{\bar q}$-dipoles are included,
 nuclear shadowing in the $\gamma A \to J/\psi A$ scattering amplitude 
is a higher twist effect~\cite{Frankfurt:2002kd}.
%vg2014 -----------------------
At the same time, the dipole approach describes reasonably well the RHIC UPC data corresponding to lower
energies (larger values of $x$), where the effect of nuclear shadowing is rather small, 
see Fig.~\ref{phenix} and its discussion in Sect.~\ref{subsec:neutron_emission}.
%------------------------------ 

In contrast to the dipole formalism, one can use the leading twist framework of QCD factorization
theorems, which enables one to apply Eq.~(\ref{eq:sigma_ryskin}) directly to nuclear targets.
Applying Eq.~(\ref{eq:sigma_ryskin}) to the nucleus and proton targets, we obtain the following 
expression for the $t$-integrated cross section of coherent $J/\psi$ photoproduction on nuclei at high energy:
\begin{equation}
\sigma_{\gamma A\rightarrow J/\psi A}^{\rm LTA}(W_{\gamma p})=\frac{d\sigma_{\gamma p\rightarrow J/\psi p}^{\rm pQCD}(W_{\gamma p},t=0)}{dt} \left[\frac{G_A(x,\mu^2)}{AG_N(x,\mu^2)}\right]^2 \Phi_A(t_{\rm min}) \,,
\label{eq:sigma_pQCD}
\end{equation}
where $G_A(x,\mu^2)$ and $G_N(x,\mu^2)$ are the gluon distributions in a nucleus and the free proton, 
respectively; 
%vg2014
$d\sigma_{\gamma p\rightarrow J/\psi p}^{\rm pQCD}(W_{\gamma p},t=0)/dt$ is the perturbative QCD (pQCD)
cross section on the proton calculated using Eq.~(\ref{eq:sigma_ryskin}).
Thus, the nuclear modification (suppression) of 
$\sigma_{\gamma A\rightarrow J/\psi A}^{\rm LTA}(W_{\gamma p})$ is given by the factor of
$R=G_A(x,\mu^2)/[AG_N(x,\mu^2)] < 1$ quantifying the amount of nuclear gluon shadowing at small $x$.

The leading twist theory of nuclear shadowing~\cite{Frankfurt:2011cs} is based on
the space--time picture of the strong interaction 
at high energies, 
the generalization of the Gribov--Glauber theory
of nuclear shadowing in soft hadron--nucleus scattering~\cite{Glauber:1955qq,Gribov_1969} 
to hard processes with nuclei, and the QCD collinear 
factorization theorems for the total and diffractive cross sections of deep inelastic
scattering (DIS). The approach   
allows one to make
predictions for the leading twist shadowing correction to nuclear parton distributions (nPDFs), structure 
functions and cross sections, 
which
are given as a series in the number of simultaneous interactions
with the target nucleons (the multiple scattering series). 
The structure of each term in the series is unambiguously given by the Gribov--Glauber 
theory supplemented by Abramovsky--Gribov--Kancheli (AGK) cutting rules~\cite{Abramovsky:1973fm} and the QCD
factorization theorems. 
%vg removed
% -------------------------------
%The nuclear shadowing effect arises as a result of 
%vg destructive quantum-mechanical interference among different terms of the multiple scattering series.
% ------------------------------- 

\begin{figure}[h]
\centering
\epsfig{file=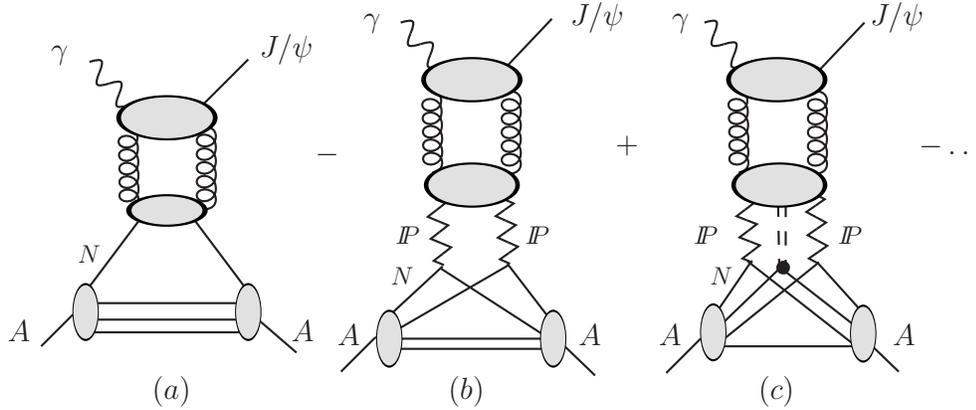,scale=0.8}
\caption{The multiple scattering series for the $\gamma A \to J/\psi A$ scattering amplitude in the leading twist theory of nuclear shadowing: (a) the impulse approximation,  (b) the double scattering, (c) the interaction with three nucleons of the target.}
\label{fig:Series_Jpsi_coh}
\end{figure}

In the graphic form, the multiple scattering series for the $\gamma A \to J/\psi A$ scattering amplitude 
in the leading twist theory of nuclear shadowing is shown in Fig.~\ref{fig:Series_Jpsi_coh},
where graph $a$ is the impulse approximation, graph $b$ corresponds to double scattering
(the simultaneous interaction of the probe with two nucleons of the target), and graph $c$ corresponds to
the interaction with three nucleons of the target.

The multiple scattering series of Fig.~\ref{fig:Series_Jpsi_coh} can be summed as follows. 
The Gribov result on the inelastic shadowing correction in hadron--nucleus
scattering
can be conveniently implemented using the formalism of cross section 
fluctuations~\cite{Good_1960}.
In this approach, the interaction of a high-energy projectile with a nucleus is a two-step process. 
First, long before the target,
the projectile fluctuates into different configurations interacting with a hadronic target 
with different cross sections $\sigma$ characterized by the distribution over cross sections $P(\sigma)$. 
Second, these fluctuations interact with the nucleus. The corresponding cross section is 
calculated separately for each fluctuation (for individual $\sigma$)
using the Glauber method
and then averaged with $P(\sigma)$, 
for details and references, see~\cite{Frankfurt:2011cs}.
In particular, for the 
$\gamma A \to J/\psi A$ scattering amplitude, we obtain:
%\begin{samepage}
\begin{eqnarray}
&&{\cal M}_{\gamma A \to J/\psi A}(t=0) \nonumber\\
&=& \varkappa \int_0^{\infty} d \sigma P(\sigma)
\int d^2 \vec b \left[\frac{\sigma\, T_A(\vec b)}{2}-\frac{\sigma^2 \, T_A^2(\vec b)}{2^2\, 2!}+
\frac{\sigma^3\, T_A^3(\vec b)}{2^3\, 3!}- \dots \right] \nonumber\\
&=& \varkappa
\int d^2 \vec b \left[\frac{\langle \sigma \rangle \, T_A(\vec b)}{2}-\frac{\langle \sigma^2 \rangle \, T_A^2(\vec b)}{2^2\, 2!}+
\frac{\langle \sigma^3 \rangle \, T_A^3(\vec b)}{2^3\, 3!}- \dots \right] \nonumber\\
&=& \varkappa A \frac{\langle \sigma \rangle}{2} \left[1
-\frac{2}{A}
\int d^2 \vec b \left(\frac{\langle \sigma^2 \rangle}{\langle \sigma \rangle}
\frac{T_A^2(\vec b)}{2^2\, 2!}-
\frac{\langle \sigma^2 \rangle}{\langle \sigma \rangle}\frac{\langle \sigma^3 \rangle}{\langle \sigma^2 \rangle}
 \frac{T_A^3(\vec b)}{2^3\, 3!}+\dots \right) \right] \,,
 \label{sfac}
\end{eqnarray}
%\end{samepage}
where $\langle \sigma^N \rangle =\int  d\sigma P(\sigma )\sigma^N$. The factor of
 $\varkappa$ contains the factors associated with the overlap of the photon and $J/\psi$ wave functions;
its value is determined by the elementary $\gamma p \to J/\psi p$ cross section:
$d\sigma_{\gamma p \to J/\psi p}^{\rm pQCD}(t=0)/dt=\varkappa^2 \langle \sigma \rangle^2/(16 \pi)$.

The first term in Eq.~(\ref{sfac}) describes photoproduction of $J/\psi$ on a single nucleon
and, hence, 
is proportional to the number of nucleons $A$; it  
is the impulse approximation corresponding to graph $a$ in Fig.~\ref{fig:Series_Jpsi_coh}.

The second term in Eq.~(\ref{sfac}) corresponds to the simultaneous interaction of the hard probe 
with two nucleons of the target nucleus and gives  the leading contribution
to the shadowing correction; 
%vg2014 to the impulse approximation;
this term 
corresponds to graph $b$ in  Fig.~\ref{fig:Series_Jpsi_coh}.
According to the Gribov--Glauber theory of nuclear shadowing 
supplemented by the collinear factorization theorem for
hard diffraction in deep inelastic scattering (DIS)~\cite{Collins:1997sr}, 
this contribution is unambiguously expressed in terms 
of elementary diffraction, notably, in terms of the diffractive gluon distribution of the proton 
$G_N^{D(3)}$~\cite{Aktas:2006hy}.
The corresponding interaction cross section is
$\sigma_2(x,\mu^2)$:
\begin{equation}
\frac{\langle \sigma^2 \rangle}{\langle \sigma \rangle} \equiv \sigma_2(x,\mu^2)=\frac{16 \pi B_{\rm diff}}{(1+\eta^2) xG_N(x,\mu^2)} \int_x^{0.1} dx_{\Pomeron} 
\beta G_N^{D(3)}(\beta,\mu^2,x_{\Pomeron}) \,,
\label{eq:sigma2}
\end{equation}
where $B_{\rm diff} \approx 6$ GeV$^{-2}$ is the slope of the $t$ dependence of the diffractive cross section;
$\eta \approx 0.17$ is the ratio of the real to the imaginary parts of the diffractive 
scattering amplitude; the diffractive parton distribution  $G_N^{D(3)}(\beta,\mu^2,x_{\Pomeron})$ 
depends on the two light-cone fractions: $x_{\Pomeron} \approx (M_X^2+\mu^2)/W_{\gamma p}^2$ is the nucleon momentum
fraction carried by the diffractive exchange presented by a zigzag line in Fig.~\ref{fig:Series_Jpsi_coh} 
($M_X$ is the invariant mass of the intermediate diffractive state)
and
$\beta=x/x_{\Pomeron}$ is the diffractive exchange (``Pomeron'') 
momentum fraction carried by the active parton.

The structure of the interaction with three and more nucleons of the target 
(graph $c$ in Fig.~\ref{fig:Series_Jpsi_coh} and higher terms that we do not show)
presents extension of that of graph $b$: in the interaction with $N$ nucleons of the target,
the hard probe diffractively scatters off two nucleons 
of the target and the produced diffractive state rescatters on the remaining $N-2$ nucleons, 
which leads to its attenuation (absorption).
In particular, the third term in Eq.~(\ref{sfac}) corresponds to the simultaneous
interaction of the hard probe with three nucleons of the target; its contribution  
corresponds to graph $c$ in  Fig.~\ref{fig:Series_Jpsi_coh}.
This contribution cannot in general 
be expressed only in terms of diffractive distributions of the proton and needs to be modeled. 
Since the cross section of hard diffraction in $ep$ DIS
exhibits the $W_{\gamma p}$ dependence typical for soft 
processes, it appears plausible to model the rescattering cross section responsible for the interaction with
$N \geq  3$ nucleons (the solid circle in graph $c$ in  Fig.~\ref{fig:Series_Jpsi_coh})
using 
the formalism of cross section fluctuations.
Exactly this was assumed in Eq.~(\ref{sfac}); the corresponding effective cross section is
\begin{equation}
\frac{\langle \sigma^3 \rangle}{\langle \sigma^2 \rangle} \equiv \sigma_3 \,,
\label{eq:sigma3}
\end{equation}
where we suppressed the $x$ and $\mu^2$ dependence of $\sigma_3$ for brevity.
In practice, the $\sigma_3$ cross section is calculated using 
the distribution $P(\sigma)$ modeled 
using the dipole formalism or $P_{\pi}(\sigma)$ of the pion, see details in~\cite{Frankfurt:2011cs}.
This is reasonable at $\mu^2 \sim $ few GeV$^2$, where soft physics dominates.
For larger values of $\mu^2$ (e.g., in the case of $\Upsilon$ photoproduction), 
one can use Eq.~(\ref{eq:sigma3}) only as input at the low initial scale for the subsequent 
Dokshitzer--Gribov--Lipatov--Altarelli--Parisi (DGLAP) evolution to the desired value of $\mu^2$.

For the interaction with $N \geq 4$ nucleons (not shown in Fig.~\ref{fig:Series_Jpsi_coh}), we assume that 
the effect of cross section fluctuations is the same as for the $N=3$ term, i.e., 
$\langle \sigma^N \rangle=\langle \sigma^2 \rangle \sigma_3^{N-2}=\langle \sigma \rangle \sigma_2 \sigma_3^{N-2}$ for $N \geq 3$.

With this input, the multiple scattering series in Eq.~(\ref{sfac}) can be summed and the result presented
in the following compact form:
\begin{samepage}
\begin{eqnarray}
{\cal M}_{\gamma A \to J/\psi A}(t=0) &=& A \frac{\varkappa \langle \sigma \rangle}{2}
 \left[1-\frac{2}{A} \frac{\sigma_2}{\sigma_3^2}
\int d^2 \vec b \left(e^{-\sigma_3/2 T_A(b)}-1+\frac{\sigma_3}{2}T_A(b) \right) \right] \nonumber\\
 &=& A \frac{\varkappa \langle \sigma \rangle}{2}
 \left[1-\frac{\sigma_2}{\sigma_3}+
\frac{\sigma_2}{\sigma_3} \frac{\sigma^A_3}{A \sigma_3}\right] \,,
 \label{sfac_2}
\end{eqnarray}
\end{samepage}
where 
$\sigma_3^A = 2 \int d^2 \vec b (1-e^{-\sigma_3/2 T_A(b)})$ 
is the total hadron--nucleus cross section in the case when the total hadron--nucleon cross section is
$\sigma_3$.
Note that in Eq.~(\ref{sfac_2}) we did not take 
into account the small real part of the soft scattering amplitude corresponding to the $\sigma_3$ 
cross section, whose numerical effect is small.

Expressing the $\gamma A\rightarrow J/\psi A$ differential cross section in terms of
the amplitude of Eq.~(\ref{sfac_2}), we obtain:
\begin{eqnarray}
\sigma_{\gamma A\rightarrow J/\psi A}^{\rm LTA}(W_{\gamma p})=\frac{d\sigma_{\gamma p\rightarrow J/\psi p}^{\rm pQCD}(W_{\gamma p},t=0)}{dt} \left[1-\frac{\sigma_2}{\sigma_3}+
\frac{\sigma_2}{\sigma_3} \frac{\sigma^A_3}{A \sigma_3} \right]^2 \Phi_A(t_{\rm min})  \,.
\label{eq:sigma_pQCD_2}
\end{eqnarray}
Note that the expression in the square brackets in nothing but the
nuclear gluon shadowing ratio $R=G_A(x,\mu^2)/[AG_N(x,\mu^2)]$, 
i.e.,
Eqs.~(\ref{eq:sigma_pQCD_2}) and (\ref{eq:sigma_pQCD}) are consistent with each other.

It is important to note that unlike the case of the color dipole formalism, 
the shadowing correction in Eq.~(\ref{eq:sigma_pQCD_2})
(and also in $R$) 
is a  
leading twist quantity determined by the elementary hard diffraction in lepton--proton DIS.
In the low nuclear density limit, when the interaction with $N \geq 3$ nucleons can be neglected,
the shadowing correction is driven by the leading twist $\sigma_2$ cross section. At the low values of 
$x$, the $N \geq 3$ terms also become important; their contributions are also leading twist quantities, 
which can be summed using the $\sigma_3$ cross section.

It is convenient to characterize the suppression of 
$\sigma_{\gamma A\rightarrow J/\psi A}^{\rm LTA}(W_{\gamma p})$ due to nuclear gluon shadowing 
in terms of
the $S^{\rm LTA}_{\rm coh}(W_{\gamma p})$ ratio:
\begin{equation}
S^{\rm LTA}_{\rm coh}(W_{\gamma p})  \equiv  
\left[\frac 
{\sigma^{\rm LTA}_{\gamma A\rightarrow J/\psi A}(W_{\gamma p})} 
{\sigma^{\rm IA}_{\gamma A\rightarrow J/\psi A}(W_{\gamma p})}\right]^{1/2}
=  \left[1-\frac{\sigma_2}{\sigma_3}+
\frac{\sigma_2}{\sigma_3} \frac{\sigma^A_3}{A \sigma_3} \right] 
= R(x,\mu^2)
\,,
\label{lorat}
\end{equation}
where $\sigma^{\rm IA}_{\gamma A\rightarrow J/\psi A}(W_{\gamma p})$ is
the $\gamma A\rightarrow J/\psi A$ cross section in the impulse approximation (IA) neglecting 
all nuclear effects except for coherence:
\begin{equation}
\sigma_{\gamma A\rightarrow J/\psi A}^{\rm IA}(W_{\gamma p})=\frac{d\sigma_{\gamma p\rightarrow J/\psi p}^{\rm pQCD}(W_{\gamma p},t=0)}{dt} \Phi_A(t_{\rm min}) \,.
\label{eq:sigma_pQCD_IA}
\end{equation}
%vg2014: do not really need
%-----------------------------
%The square brackets in Eq.~(\ref{lorat}) give the nuclear suppression factor arising from multiple 
%interactions of the produced diffractive state with target nucleons: the first term is the impulse 
%approximation and the second and the third terms describe the shadowing correction 
%driven by the $\sigma_2$ cross section (for the interaction with $N=2$ nucleons) and 
%by the $\sigma_3$ cross section (for the interaction with $N \geq 3$ nucleons).  
%-----------------------------

Theoretical predictions for the suppression factor of $S^{\rm LTA}_{\rm coh}$ of Eq.~(\ref{lorat})
calculated using the leading twist theory of nuclear shadowing~\cite{Frankfurt:2011cs}
and the EPS09 global QCD fits~\cite{Eskola:2009uj} agree 
 well with the nuclear suppression 
factor obtained in the recent analysis of the ALICE data 
on coherent $J/\psi$ photoproduction in Pb-Pb UPCs
at $x\approx 10^{-2}$ and $x\approx 10^{-3}$~\cite{Guzey:2013xba,Guzey:2013qza}  
(see also Fig.~\ref{cirapalice}).

\subsection{The incoherent nuclear $J/\psi$ photoproduction cross section}
\label{subsec:incoherent_case}

Similarly to the case of coherent $J/\psi$ photoproduction on nuclei considered in 
Sect.~\ref{subsec:coherent_case}, the VMD model, the color dipole formalism
and the leading twist approximation can be used to calculate the cross section of incoherent 
nuclear $J/\psi$ photoproduction, $\sigma_{\gamma A\rightarrow J/\psi A^{\prime}}$, where 
$A^{\prime}$ denotes the final nuclear state containing products of the nuclear disintegration
($A^{\prime} \neq A$). Using completeness of the $A^{\prime}$ states, 
in the high-energy optical limit of the Glauber model, the VMD model gives~\cite{Bauer:1977iq}:
\begin{equation}  
\frac{\sigma_{\gamma A\rightarrow J/\psi A^{\prime}}^{\rm VMD}(W_{\gamma p})}{dt}=
\frac{d\sigma_{\gamma p\rightarrow J/\psi p}(W_{\gamma p})}{dt}\int d^2 \vec{b}\, T_A(b) e^{-\sigma_{J/\psi N}^{\rm in}(W_{\gamma p}) T_A(b)}
 \,,
\label{eq:sigma_incoh_VMD}
\end{equation}
where 
$\sigma_{J/\psi N}^{\rm in}=\sigma_{J/\psi N}-\sigma_{J/\psi N}^2/(16 \pi B_{J/\psi})$ is the inelastic 
$J/\psi$--nucleon cross section; $B_{J/\psi}$ is the slope of the $t$ dependence of the 
$\gamma p\rightarrow J/\psi p$ scattering amplitude. Note that 
Eq.~(\ref{eq:sigma_incoh_VMD}) is valid at not too small $|t| \neq 0$.

Equation~(\ref{eq:sigma_incoh_VMD}) has the straightforward and well-known interpretation: 
the probability of incoherent (quasielastic) photoproduction of a vector meson on a nucleus is 
given by the product of the probability of elastic scattering on a single nucleon times the probability
for the produced vector meson to survive the passage through the nucleus on its way out.
Since $\sigma_{J/\psi N}^{\rm in} \approx \sigma_{J/\psi N}$ is small, 
the nuclear suppression of $\sigma_{\gamma A\rightarrow J/\psi A^{\prime}}^{\rm VMD}$ due to 
nuclear shadowing is
also small.

In the color dipole formalism, the coherent and incoherent nuclear $J/\psi$ photoproduction cross sections
can be calculated on the same footing. The only difference is the order of averaging over the dipole sizes 
$d_t$: in the coherent case, one first averages the $\gamma A \to J/\psi A$ amplitude over $d_t$ and then squares 
the result, while in the incoherent case, one first squares the appropriate scattering amplitude and then
averages the result over $d_t$, see, e.g., Ref.~\cite{Lappi:2013am}. 
Since the relevant dipole cross section is small, similarly to the coherent case considered above, 
nuclear suppression of incoherent nuclear $J/\psi$ photoproduction cross sections is small~\cite{Lappi:2013am}.
%vg2014 rewritten
At small $x$ typical for the LHC kinematics, 
%vg $x \sim 10^{-4}-10^{-2}$,
 $x \sim 10^{-3}$ and below,
% While 
the dipole formalism predictions are subject to rather significant theoretical
uncertainties due to the choice of the model for the dipole cross section and for the $J/\psi$ wave function.
Nevertheless, one can still make the observation
%vg2014 
that the shadowing suppression of the incoherent cross section of $J/\psi$ photoproduction 
on nuclei appears to be larger than that for the coherent case.
%vg2014
%and (ii) the suppression of the incoherent nuclear $J/\psi$ 
%photoproduction cross section is smaller by approximately a factor of $1.5$ than that in the leading 
%twist approach, compare Fig.~2 of~\cite{Lappi:2013am}
%with Fig.~\ref{cirapalice} of this work.
%---------------------------------
%vg2014 new sentence
At the same time, in the RHIC kinematics corresponding to much larger values of $x$, $x \approx 10^{-2}$, 
where the effect of nuclear shadowing is small, both the dipole framework
and the leading twist approach
provide the good description of the 
PHENIX UPC data~\cite{Afanasiev:2009hy}.
%vg2014, see Fig.~\ref{phenix} in Sect.~\ref{subsec:neutron_emission}. 
Note also that at 
$x \sim 10^{-2}$, numerous versions of the dipole model correspond to the similar color dipole cross section
because the models have been fitted to the same data, see, e.g., Refs.~\cite{Motyka:2008jk,Rezaeian:2012ji}.
%vg2014
% As a result, the predicted magnitudes 
%of nuclear shadowing are similar and are also close to that predicted in the leading twist approach.
%---------------------------------------

In the leading twist formalism, the cross section of incoherent $J/\psi$ photoproduction on nuclei
can be readily calculated using the input employed in the coherent 
case~(\ref{eq:sigma_pQCD_2}). Generalizing the standard expression for the incoherent (quasielastic) nuclear 
cross section~\cite{Bauer:1977iq} to include cross section fluctuations, 
we obtain in the high-energy limit:
\newpage
  \begin{eqnarray}  
\frac{\sigma_{\gamma A\rightarrow J/\psi A^{\prime}}^{\rm LTA}(W_{\gamma p})}{dt}&=& \varkappa^2 e^{t B_{J/\psi}}\int d^2 \vec{b}\, T_A(b) 
\int d\sigma P(\sigma) \int d\sigma^{\prime} P(\sigma^{\prime}) \nonumber\\
&\times&
 \frac{\sigma \sigma^{\prime}}{16 \pi} e^{-\sigma/2 T_A(b)} e^{-\sigma^{\prime}/2 T_A(b)}e^{\sigma \sigma^{\prime}/(16 \pi B_{\rm diff}) T_A(b)}
 \,.
\label{eq:sigma_incoh_QCD}
\end{eqnarray}
Note that in Eq.~(\ref{eq:sigma_incoh_QCD}), averaging over cross section fluctuations should be
performed at the amplitude level which explains presence of two integrals over $\sigma$ and 
$\sigma^{\prime}$. Note also that in Eq.~(\ref{eq:sigma_incoh_QCD}) we assumed that the slopes of the $t$
dependence of the soft scattering amplitudes corresponding to the $\sigma_2$ and $\sigma_3$ cross sections
are equal; the numerical effect associated with the inequality of the slopes is negligibly small.

Recalling that 
$d\sigma_{\gamma p \to J/\psi p}^{\rm pQCD}(t=0)/dt=\varkappa^2 \langle \sigma \rangle^2/(16 \pi)$ and 
Eqs.~(\ref{eq:sigma2}) and (\ref{eq:sigma3}), after some algebra, 
 Eq.~(\ref{eq:sigma_incoh_QCD}) can be written in the following compact form in terms of 
the $\sigma_2$ and $\sigma_3$ cross sections:
\begin{eqnarray}
&&\frac{\sigma_{\gamma A\rightarrow J/\psi A^{\prime}}^{\rm LTA}(W_{\gamma p})}{dt} = \frac{d\sigma_{\gamma p \to J/\psi p}^{\rm pQCD}(W_{\gamma p})}{dt}
\int d^2 {\vec b}\, T_A(b) \nonumber\\
&\times&
\left[\left(1-\frac{\sigma_2}{\sigma_3}\right)^2+2\frac{\sigma_2}{\sigma_3}\left(1-\frac{\sigma_2}{\sigma_3}\right)  e^{-\sigma_3/2 T_A(b)}+\left(\frac{\sigma_2}{\sigma_3}\right)^2 e^{-\sigma_3^{\rm in} T_A(b)}
\right] \nonumber\\
& \approx & \frac{d\sigma_{\gamma p \to J/\psi p}^{\rm pQCD}(W_{\gamma p})}{dt}
\int d^2 \vec{b}\, T_A(b) 
\left[1-\frac{\sigma_2}{\sigma_3}+\frac{\sigma_2}{\sigma_3}e^{-\sigma_3/2 T_A(b)}
\right]^2
 \,, 
\label{eq:sigma_incoh_2}
\end{eqnarray}
where $\sigma_3^{\rm in}=\sigma_3 -\sigma_3^2/(16 \pi B_{\rm diff})$.
In the last line of Eq.~({\ref{eq:sigma_incoh_2}) we used $\sigma_3^{\rm in} \approx \sigma_3$; we checked
that this approximation works with high accuracy at the level of a few percent. 

The physical interpretation of Eq.~(\ref{eq:sigma_incoh_2}) is 
similar to that of Eqs.~(\ref{sfac_2}) and (\ref{lorat}): the nuclear suppression factor in 
the square brackets arises from multiple interactions of the produced diffractive state with 
nucleons of the target, which are driven by the $\sigma_2$ and $\sigma_3$ cross sections.

A comparison of Eqs.~(\ref{eq:sigma_incoh_2}) and (\ref{eq:sigma_pQCD_2}) shows that 
in the leading twist approximation (LTA), nuclear suppression in both
coherent and incoherent photoproduction is determined by the
same quantities: $\sigma_{2}$ and $\sigma_{3}$
[see Eqs.~(\ref{lorat}) and (\ref{eq:sigma_incoh_3})].
 The $\sigma_{2}$ cross section
is a model-independent quantity whose magnitude and 
$x$ dependence are fixed by the experimentally measured 
diffractive parton distributions, inclusive gluon distributions and DGLAP evolution equations,
see Eq.~(\ref{eq:sigma2}).
The $\sigma_{3}$ cross section is a model-dependent quantity of the LTA approach, whose
value is constrained using the formalism of cross section fluctuations. 
In general, $\sigma_{3}\geq \sigma_{2}$ [see Eq.~(\ref{eq:sigma3})]; the lower limit on
the value of $\sigma_{3}$, $\sigma_{3}= \sigma_{2}$, corresponds to the upper limit on the
predicted nuclear shadowing.

Equation~(\ref{eq:sigma_incoh_2}) defines the shadowing suppression factor for incoherent nuclear
$J/\psi$ photoproduction, $S_{\rm incoh}$:
\begin{eqnarray}
S_{\rm incoh}^{\rm LTA}(W_{\gamma p}) &\equiv& \frac{d\sigma_{\gamma A \to J/\psi A^{\prime}}^{\rm LTA}(W_{\gamma p})/dt}{A d\sigma_{\gamma p \to J/\psi p}^{\rm pQCD}(W_{\gamma p})/dt} \nonumber\\
&=& \frac{1}{A}
\int d^2 \vec{b}\, T_A(b) \left[1-\frac{\sigma_2}{\sigma_3}+\frac{\sigma_2}{\sigma_3}e^{-\sigma_3/2 T_A(b)} \right]^2  \,. 
\label{eq:sigma_incoh_3}
\end{eqnarray}
Note that Eqs.~(\ref{eq:sigma_incoh_2}) and (\ref{eq:sigma_incoh_3}) are valid at not too small $|t| \neq 0$. 

%vg2014
Figure~\ref{fig:S_incoh_fluct} presents our predictions for $S_{\rm incoh}^{\rm LTA}$ as a function of
$x=M_{J/\psi}^2/W_{\gamma p}^2$ for $^{208}$Pb at $\mu^2=3$ GeV$^2$. The shaded band represents the theoretical
uncertainty of our predictions associated with the range of possible values of the $\sigma_3$ 
cross section~\cite{Frankfurt:2011cs}.

\begin{figure}[t]
\centering
\epsfig{file=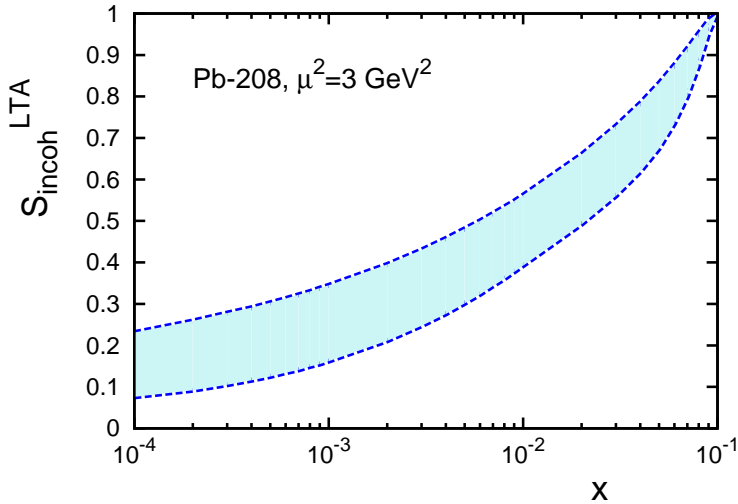,scale=1.3}
\caption{The shadowing suppression factor for incoherent nuclear
$J/\psi$ photoproduction, $S_{\rm incoh}^{\rm LTA}$, as 
a function of $x=M_{J/\psi}^2/W_{\gamma p}^2$ for $^{208}$Pb at $\mu^2=3$ GeV$^2$.
The shaded band is the theoretical uncrertainty of our LTA approach.
}
\label{fig:S_incoh_fluct}
\end{figure}

%vg2014 rewritten
% --------
%One should note that since both suppression factors of 
%$S^{\rm LTA}_{\rm coh}$~(\ref{lorat}) and $S_{\rm incoh}^{\rm LTA}$~(\ref{eq:sigma_incoh_3}) 
%are determined by the
%essentially soft physics, based on the Glauber model calculations of 
%the total and inelastic hadron--nucleus cross sections, we expect that numerically
%$S_{\rm incoh}^{\rm LTA} < (S^{\rm LTA}_{\rm coh})^2$. 
%This turns out to be also the case in the leading twist approximation, see Fig.~\ref{cirapalice}.
% -------------

One should note that since both suppression factors of 
$S^{\rm LTA}_{\rm coh}$~(\ref{lorat}) and $S_{\rm incoh}^{\rm LTA}$~(\ref{eq:sigma_incoh_3}) 
are determined by the essentially soft physics, we expect them to be numerically of a 
a similar magnitude, with $S_{\rm incoh}^{\rm LTA}$ being somewhat smaller than $(S^{\rm LTA}_{\rm coh})^2$.
Indeed, at $x=10^{-3}$ corresponding to $y \approx 0$ for Pb-Pb UPCs at $\sqrt{s_{NN}}=2.76$ GeV, we obtain
$S_{\rm incoh}^{\rm LTA}=0.16-0.35$ from Fig.~\ref{fig:S_incoh_fluct} and 
$(S_{\rm coh}^{\rm LTA})^2=0.35-0.43$ from Fig.~3 of Ref.~\cite{Guzey:2013qza}.

\section{Photoproduction of $J/\psi$ in P\MakeLowercase{b}-P\MakeLowercase{b} UPC\MakeLowercase{s} at the LHC}
\label{sec:UPCs}

\subsection{Coherent and incoherent cases}
\label{subsec:upc_coherent}

A high energy nucleus--nucleus ultraperipheral collision takes place when the colliding ions pass each other
at the distance $|{\vec b}|$ in the transverse plane (impact parameter) exceeding the sum of the nucleus radii, 
 $|\vec b|>(2 - 3)R_A$, where $R_A$ is the nuclear radius (the UPC physics is reviewed in~\cite{Baltz:2007kq}).
In this case, the strong interaction between the nuclei is suppressed and 
they interact electromagnetically via emission of quasi-real photons. Thus, 
nucleus--nucleus UPCs offer a possibility to probe very high energy photon--nucleus scattering and, 
in particular, photoproduction of $J/\psi$ on nuclei. 
The corresponding cross section has the following form:
\begin{eqnarray}
 \frac{d \sigma_{AA\to AA^{\prime}J/\psi}(y)}{dy} 
=N_{\gamma/A}(y)\sigma_{\gamma A\to J/\psi A^{\prime}}(y)+
N_{\gamma/A}(-y)\sigma_{\gamma A\to J/\psi A^{\prime}}(-y) \,,
\label{csupc}
\end{eqnarray}
where $N_{\gamma/A}(y)=\omega dN_{\gamma/A}(\omega)/d\omega$ is the photon flux; 
$y = \ln(2\omega/M_{J/\psi})$ is the $J/\psi$ rapidity, where $\omega$ is the photon 
energy and $M_{J/\psi}$ is the mass of $J/\psi$;
$\sigma_{\gamma A\to J/\psi A^{\prime}}$ is the nuclear $J/\psi$ photoproduction cross section
(see Sect.~\ref{sec:shadowing}).
Note that Eq.~(\ref{csupc}) includes both the case of coherent scattering without the nuclear breakup 
($A^{\prime}=A$) and the case of incoherent (quasielastic) 
scattering when the final nucleus dissociates ($A^{\prime} \neq A$).

The photon flux at large impact parameters $b=|{\vec b}| > 2R_A$
emitted by a fast-moving nucleus, $N_{\gamma /A}(\omega)$, 
can be approximated very well by the following 
simple semi-classical expression for the flux of equivalent 
photons produced by a point-like particle with the electric charge $Z$: 
\begin{equation}
N_{\gamma /A}(\omega)=\frac{2Z^2\alpha_{\rm e.m.}}{\pi} 
\int \limits_{2R_A}^{\infty} db\,
{X^2\over b} \left[K^2_1(X)+\frac {1} {\gamma^2_L} K^2_0(X)\right] \,,
\label{plflux}
\end{equation}
where $\alpha_{\rm e.m.}$ is the fine-structure constant; 
 $K_0(X)$ and $K_1(X)$ are modified Bessel functions;
$X=b \omega/\gamma_L$,  where $\gamma_L$ is the nucleus Lorentz factor.
The strong interactions between the
colliding nuclei is suppressed by the requirement that  $b>2R_A$.
Experimentally this corresponds to the selection of events with 
only two leptons from the  
$J/\psi$ decay and otherwise no 
charged particles in the whole rapidity range covered by the detector.

The presence of two terms in Eq.~(\ref{csupc}) reflects  
the fact that each nucleus can radiate a photon as well as can serve as a target. 
In the case of symmetric UPCs (e.g., in the case of Pb-Pb UPCs at the LHC
and Au-Au UPCs at RHIC), 
at a fixed value of the $J/\psi$ rapidity $y \neq 0$, 
Eq.~(\ref{csupc}) contains two contributions: one corresponding to the interaction of
high-energy photons with a nucleus and another corresponding to the interaction of
low-energy photons with a nucleus.

In the coherent case, 
the separation 
of these overlapping contributions is not an easy problem since a priory one cannot say 
in the interaction with which of the two nuclei $J/\psi$ was produced.
 As a result,
the photoproduction cross section $\sigma_{\gamma A\to J/\psi A}$
can be 
unambiguously
extracted from the measured $d\sigma_{AA\to AA J/\psi}(y)/dy$ cross section only
in the following two cases: (i)
at $y=0$, where the two photon energies are equal and, hence, both terms in Eq.~(\ref{csupc}) contribute
equally, and (ii) in the rapidity range where one of the contributions strongly dominates.
Exactly this situation is realized in the  ALICE experiment~\cite{Abbas:2013oua,Abelev:2012ba}:
the $d\sigma_{AA\to AA J/\psi}(y)/dy$ cross section was measured (i) for  $-1<y<1$, which allowed one
to extract $\sigma_{\gamma A\to J/\psi A}(W_{\gamma p})$ at 
$W_{\gamma p}\approx 100$ GeV corresponding to the gluon momentum fraction of $x\approx 10^{-3}$
($W_{\gamma p}$ is the $\gamma$--nucleus center-of-mass energy per nucleon)
and (ii) for $-4<y<-2.5$, where the low-energy photon contribution
dominates (more than $95\%$), which probes the nuclear gluon density at $x\approx 10^{-2}$.

As we already explained in the Introduction, 
the nuclear suppression factor for coherent nuclear $J/\psi$ photoproduction
determined from the corresponding UPC cross section measured by the ALICE 
collaboration~\cite{Guzey:2013xba,Guzey:2013qza} compares favorably with the theoretical models
predicting large nuclear gluon shadowing,
notably, with the leading twist approximation (LTA)~\cite{Frankfurt:2011cs} and with the  
EPS09~\cite{Eskola:2009uj} result.
This is illustrated in Fig.~\ref{cirapalice}, where the ALICE data on 
the coherent $d \sigma_{AA\to AAJ/\psi}(y)/dy$ cross section at the central and forward values of the rapidity $|y|$ are
compared to the LTA predictions combined with the CTEQ6L1 gluon parameterization~\cite{Pumplin:2002vw} at 
$\mu^2=3$ GeV$^2$. One can see from from Fig.~\ref{cirapalice} that
the theoretical calculations, which are made using Eqs.~(\ref{eq:sigma_pQCD_2}) and (\ref{csupc}), 
describe the data well.
%vg2014 (the red shaded band represents
%the theoretical uncertainty of the LTA predictions).

\begin{figure}[h]
\centering
\epsfig{file=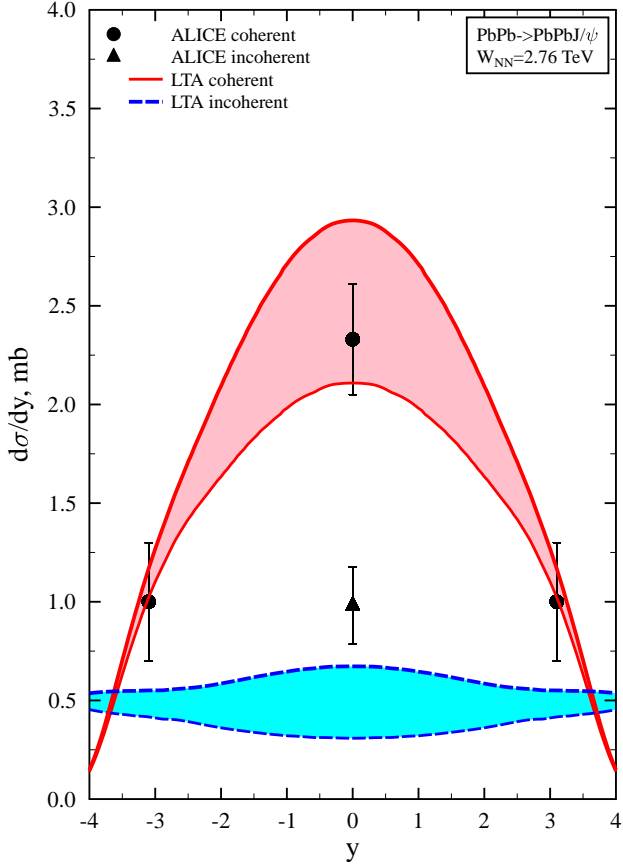,scale=0.5, trim= 0 50 0 100}
\caption{The coherent $d \sigma_{AA\to AAJ/\psi}(y)/dy$ and incoherent $d \sigma_{AA\to AA^{\prime}J/\psi}(y)/dy$ 
cross sections as functions of the $J/\psi$ rapidity $y$ at $\sqrt{s}=2.76$ GeV. The 
ALICE data~\cite{Abbas:2013oua,Abelev:2012ba} is compared to the LTA theoretical predictions; the bands span the uncertainty of the
theoretical predictions.}
\label{cirapalice}
\end{figure}

In the same figure, the LTA predictions for the incoherent $d \sigma_{AA\to AA^{\prime}J/\psi}(y)/dy$ cross section
made using  Eqs.~(\ref{eq:sigma_incoh_2}) and (\ref{csupc}) 
are compared to the {\mbox ALICE} data point at 
$|y| \approx 0$~\cite{Abbas:2013oua}. 
One can see from the comparison that the LTA 
predicts the magnitude of suppression due to nuclear gluon shadowing exceeding the one seen
in the data by approximately a factor of $1.5$.

%vg2014
The shaded bands in Fig.~\ref{cirapalice} represent the dominant theoretical uncertainty of the 
LTA predictions associated with the uncertainty in the value of the $\sigma_3$ cross section, 
which in turn results in the uncertainty of the LTA predictions for nuclear parton 
distributions~\cite{Frankfurt:2011cs}. 
The uncertainty associated with the value of the hard scale $\mu^2$, which was studied in~\cite{Guzey:2013qza}, 
is much smaller and has been safely neglected.

%vg2014 rewritten to expand and soften the statement that the ALICE data necesseraly contains
% the nucleon diffraction contribution
Note that in our calculations, we consider 
quasielastic scattering and do not take into account the
incoherent contribution associated with the nucleon dissociation 
$\gamma +N \to J/\psi +Y$~\cite{Frankfurt:2008er}. 
We explained in the Introduction that this process could potentially contribute to the 
inelastic final state and, thus, affect the ALICE extraction of the incoherent 
 $d \sigma_{AA\to AA^{\prime}J/\psi}(y)/dy$ cross section~\cite{Abbas:2013oua}
%which is included in the ALICE $d \sigma_{AA\to AA^{\prime}J/\psi}(y)/dy$ data point~\cite{Abbas:2013oua}
due to the fact that the ALICE detector does not cover the full range 
of the rapidity $y$.
%vg2014
While the calculation of the $\gamma+A \to J/\psi+Y+(A-1)^{\ast}$ contribution 
requires a a separate publication, one can still make several qualitative observations.
First, 
this contribution is expected to have approximately the same $A$ dependence as 
that in Eq.~(\ref{eq:sigma_incoh_2}) (it is proportional to $A$ in the impulse approximation).
Second, the magnitude of this contribution is expected to be sizable:
$(d\sigma_{\gamma N \to J/\psi  Y}/dt)/(d\sigma_{\gamma N \to J/\psi  N}/dt) \approx 0.15$ at $t \approx 0$ and 
increases with an increase of $|t|$ when the $d\sigma_{\gamma N \to J/\psi  Y}/dt$ cross section
 becomes progressively more important and 
eventually exceeds that of the elastic $\gamma + N \to  J/\psi +N$ process; 
$\sigma_{\gamma N \to J/\psi  Y}/\sigma_{\gamma N \to J/\psi  N} \approx 0.8$ for the $t$-integrated cross sections and 
for $M_Y < 10$ GeV ($M_Y$ is the invariant mass of the proton-dissociative system $Y$)~\cite{Alexa:2013xxa}.
It would be desirable to perform an additional analysis of the ALICE data~\cite{Abbas:2013oua}
by assuming that the $\gamma +N \to J/\psi +N$ and $\gamma +N \to J/\psi +Y$
contributions to incoherent nuclear $J/\psi$ photoproduction have different
slopes of the $t$ dependence, which would enable one to experimentally estimate 
the contribution of the nucleon dissociation 
and, thus, will enable a direct comparison of the data with predictions of Eq.~(\ref{eq:sigma_incoh_2}).
In addition, it is likely that due to the interaction of the system $Y$ with the nucleus,
nucleon dissociation will lead to
a larger number of neutrons originating from the nucleus dissociation. Finally,
the study of neutron production in the quasielastic $\gamma A \to J/\psi A^{\prime}$ process
at $|t| \ge 1$ GeV$^2$, where the $\gamma +N \to J/\psi +Y$ contribution 
dominates,
may be interesting for understanding of the formation time in diffraction.

\subsection{UPCs accompanied by neutron emission}
\label{subsec:neutron_emission}

Besides ALICE,
the ATLAS and CMS detectors at the LHC are capable to measure UPC production
of $J/\psi$ in the $-2.5<y<2.5$ range of rapidity. 
While for central rapidities,  
the interpretation of the corresponding
measurements is unambiguous, it is difficult to disentangle 
the high-photon-energy and low-photon-energy contributions to
$d \sigma_{AA\to AAJ/\psi}(y)/dy$ for 
non-central values of $y$
 and, thus, to 
access the small-$x$ region that we are interested in.
In particular,
according to the estimates of~\cite{Guzey:2013xba,Guzey:2013qza}, for $1.5<|y|<2.5$ in Pb-Pb UPCs at 2.76 TeV, 
the $d \sigma_{AA\to AAJ/\psi}(y)/dy$ rapidity distribution in the coherent case 
is 
exceedingly dominated (by the factor of four) by  the low-photon-energy contribution corresponding to
$10^{-2} > x > 5 \times 10^{-3}$ of the probed nuclear gluons.
The high-photon-energy contribution is suppressed by the much lower photon flux and the larger 
nuclear gluon shadowing.
Hence, it is rather difficult to 
extract 
the high energy nuclear coherent $J/\psi$ photoproduction cross section from the UPC data and, hence, to 
probe the nuclear gluon distribution 
around $x\approx 10^{-4}$.

The method 
to overcome this difficulty was suggested in~\cite{Rebyakova:2011vf}. 
It is based on the observation of~\cite{Baltz:2002pp} that
 coherent photoproduction of vector mesons in heavy ion UPCs 
can be accompanied by
additional photon exchanges which lead to electromagnetic excitation
of one or both nuclei with the subsequent 
neutron emission.
These neutrons will have the 
energy close to that of the colliding beams and can be detected 
by zero degree calorimeters placed at large distances on both sides 
of the detectors. 
With the additional requirement to have in the final state only
two muons from the $J/\psi$ decay (in addition to the neutrons) and the large rapidity gap, 
i.e., by requiring the absence 
of any other charged particle in the whole 
range of $y$ covered by the detector system,
the strong interaction of the colliding nuclei should be suppressed.

To calculate the cross section of
%vg coherent 
photoproduction of vector mesons in UPCs accompanied by the 
additional electromagnetic excitation of the colliding nuclei followed by their
subsequent 
neutron emission, 
we use the model
developed in~\cite{Baltz:2002pp}.
This approach is justified by the success of the calculations of~\cite{pshenichnov}
describing very well the ALICE data on electromagnetic dissociation in Pb-Pb UPCs~\cite{ALICE:2012aa}.
The model is based on the assumption that an additional photon exchange does not
destroy coherence of the photoproduction process but influences  
the impact parameter of the ultraperipheral collision. The latter is taken into account
by the modification of the flux of the photons participating in photoproduction: 
\begin{equation}
 N_{\gamma /A}^i(\omega)= \int \limits_{2R_A}^{\infty} d^2 {\vec b}\,  
 N_{\gamma /A}(\omega ,\vec b)P_{i}(\vec b) \,,
 \label{flux}
\end{equation}
where the impact parameter dependent factor of $P_{i}(\vec b)$ 
takes into account different channels of 
the nuclear decay by the neutron emission ($i=$~0n0n, 0nXn, XnXn, $\dots$).
In particular, the 0n0n-channel corresponds to the selection of events without additional
electromagnetic dissociation with the nucleus
neutron decay; the 0nXn-channel corresponds to one-side excitation
with the nucleus
neutron decay of only one of the colliding nuclei; the XnXn-channel corresponds to mutual electromagnetic
dissociation with both excited nuclei decaying by neutron emission.
To obtain a rough estimate of the size of the effect,
each additional photon exchange in Pb-Pb UPCs leads to the suppression
of the cross section by the factor of $Z^{2}{\alpha}^{2}_{\rm e.m.} \approx 0.3-0.4$.

Based on the assumption that an additional photon exchange influences 
only the flux of photons, see Eq.~(\ref{flux}), one can try to separate the low-energy ($\omega _1$)
and high-energy ($\omega _2$) $J/\psi$  photoproduction contributions 
in the rapidity spectra measured in heavy ion UPCs.
To this end, 
it is necessary \cite{Rebyakova:2011vf} to measure rapidity distributions 
of vector meson production in any two channels, e.g., 0nXn and XnXn,
and using the calculated photon fluxes for these channels, to solve
simple equations at a fixed value of the rapidity $y$.
For the coherent case, one then obtains:
\begin{eqnarray}
d\sigma^{0nXn}/dy &=& 
N^{0nXn}_{\gamma} (\omega_1 ) \sigma_{\gamma A \rightarrow J/\psi A} (\omega_1 )+ 
N^{0nXn}_{\gamma} (\omega_2 ) \sigma_{\gamma A \rightarrow J/\psi A} (\omega_2 ) \,, \nonumber\\
d\sigma^{XnXn}/dy &=&
N^{XnXn}_{\gamma} (\omega_1 ) \sigma_{\gamma A \rightarrow J/\psi A} (\omega_1 )+
N^{XnXn}_{\gamma} (\omega_2 ) \sigma_{\gamma A \rightarrow J/\psi A} (\omega_2 ) \,.
\label{eq:disentangling}
\end{eqnarray}
Since the photon fluxes $N^{0nXn}_{\gamma}$ and $N^{XnXn}_{\gamma}$
can be calculated with good accuracy, measurements
of different channels will allow one to study nuclear gluon shadowing in a wide
range of $x$.
According to our calculations presented and discussed below, 
the relative contributions of low-photon-energy and high-photon-energy $J/\psi$
photoproduction in these channels are strongly different: while in the 0nXn-channel they
are almost equal, the high-energy photoproduction dominates in the XnXn-channel.

Production of forward neutrons in quasielastic incoherent 
photoproduction of $J/\psi$ in heavy ion UPCs with the nuclear breakup
has been considered in~\cite{Strikman:2005ze}.
Since 
in this case
the momentum transfer
in elastic
$J/\psi$ photoproduction on the nucleon 
can be as large as $|t|=1$ GeV$^2$, this target
nucleon escaping from the nucleus participates in additional
quasielastic rescattering. The average excitation energy
of a heavy nucleus in the one-nucleon removal process is about $20-25$ MeV, which
is much higher than the separation energy of $7-8$ MeV of one 
neutron.
It was shown in~\cite{Strikman:2005ze}
 that in incoherent $J/\psi$
photoproduction in heavy ion UPCs,
the residual nucleus will decay emitting one or more neutrons
with the probability of about $85\%$.
Therefore, imposing the constraint that no neutrons are emitted, i.e., considering the
0n0n-channel, one can almost completely (at the level of $10-15$\%) suppress the incoherent contribution.

As we mentioned in the end of Sect.~\ref{subsec:upc_coherent}}, the contribution of nucleon dissociation 
becomes important/dominant with an increase of the transverse momentum of $J/\psi$.
This process should lead to at least as many neutrons as the quasielastic process.
Therefore,  the $\gamma +N \to J/\psi +N$ and 
$\gamma +N \to J/\psi +Y$ contributions should be either separated experimentally or the latter should be
included in the theoretical calculation of the $\gamma A \to J/\psi A^{\prime}$ cross section.
The procedure for the extraction of the high-photon-energy contribution that we discuss below 
involves the  
use of the different $p_t$ dependences of the $\gamma +N \to J/\psi +N$ and 
$\gamma +N \to J/\psi +Y$ cross sections, which
allows one to separate their contributions.
We also note in passing that the study of the neutron multiplicity at $p_t \ge 0.8$ GeV, 
where the process of nucleon dissociation dominates, would produce for the first time 
 information about the space--time  formation 
of hadrons in the diffractive processes like $\gamma +N \to J/\psi +Y$.

\begin{figure}[h]
\centering
\epsfig{file=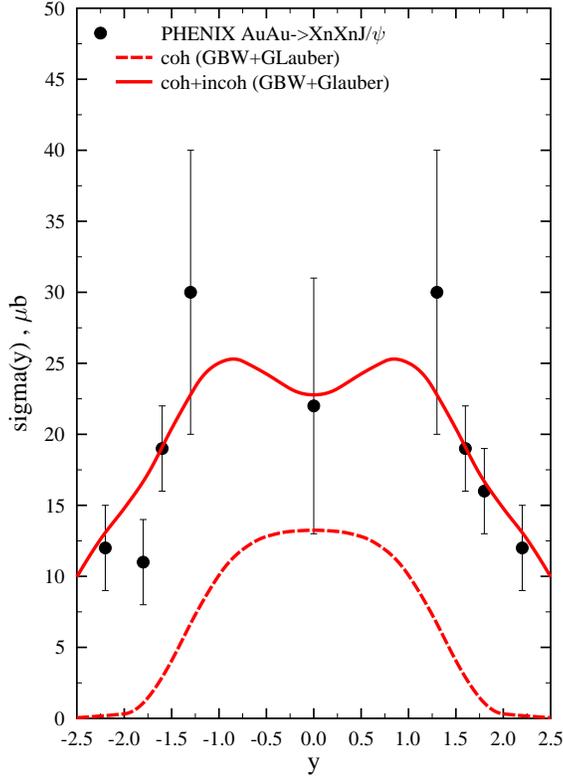,scale=0.45, trim= 0 50 0 100}
\caption{The rapidity distribution of $J/\psi$ photoproduction in Au-Au UPCs with neutron emission in the 
XnXn-channel at $\sqrt{s}=200$ GeV. The PHENIX data~\cite{takahara} is compared to the theoretical
prediction made 
%vgin the color dipole framework.
using the simple dipole model described in the text (the curves labeled ``GBW+Glauber'').
The upper solid curve corresponds to the sum of the coherent and incoherent contributions; 
the lower dashed curve corresponds to the purely coherent contribution. 
}
\label{phenix}
\end{figure}

The recent PHENIX data on $J/\psi$ photoproduction
in ultraperipheral Au-Au collisions at 200 GeV
accompanied by  neutrons detected in both ZDCs~\cite{Afanasiev:2009hy,takahara} 
gives an opportunity to check 
main assumptions about the sources of forward neutrons in 
coherent and incoherent $J/\psi$ photoproduction in 
heavy ion UPCs at high energies. Indeed, since the nuclear gluon shadowing 
in these processes is small at RHIC energies (the shadowing effect leads to an approximately 20\%
reduction at the cross section level), 
 the coherent and incoherent photoproduction cross
sections can be calculated using 
%vg2014 -------
either the dipole model or the leading twist approximation;
at $x \sim 10^{-2}$ corresponding to the RHIC kinematics, the dipole model and LTA predictions 
largely converge. In this work, we use a simple version of the dipole model which employs
the standard Glauber expressions [Eqs.~(\ref{sigglc}) and 
(\ref{eq:sigma_incoh_VMD})]
 with the elementary $\sigma_{J/\psi N}$ cross section approximated by
the phenomenological Golec-Biernat--Wusthoff dipole 
cross section~\cite{GolecBiernat:1998js}.
The resulting predictions are labeled ``GBW+Glauber'' in Fig.~\ref{phenix}.

It is worth noting here that for the discussed kinematics, 
the results for the dipole--nucleon cross section 
obtained in different dipole models are rather close 
since they are well constrained by the DIS data for these 
%vg2014
energies~\cite{Motyka:2008jk,Rezaeian:2012ji}.
The theoretical 
uncertainty is much smaller than the PHENIX experimental errors and, hence, it is not shown
in Fig.~\ref{phenix}.
%-------------------
Note also that in the discussed model, the nuclear shadowing  effect is 
driven by the 
$\sigma_{c\bar c N}$ dipole cross section and, hence, shadowing is suppressed 
(it is a higher twist effect)  
for the dipoles of such a small size, see the discussion above.

Selection of events with two-side neutron detection 
means that 
in the XnXn-channel,
coherent production should 
take into account 
the mutual electromagnetic dissociation that requires
at least two additional photon exchanges.  
In contrast, 
according to the predictions of~\cite{Strikman:2005ze}, incoherent production
is associated with 
excitation and neutron decay of only the target nucleus. Hence, 
in this case,  detection of the XnXn-channel requires only one photon 
exchange to excite the nucleus,
which serves as a source of the photon flux
 in the process of incoherent production.

Figure~\ref{phenix} presents a comparison of the PHENIX results for the XnXn-channel with our
theoretical calculations using the simple dipole 
described above. 
%vg
%vg2014 rewritten
% -----------------------
%Note that as we discussed above, 
%One can see from the figure that within rather
%large experimental errors, the agreement between the data and theory is reasonable.  
%Hence, the results presented
%in Fig.~\ref{phenix} seem to confirm the prediction of~\cite{Strikman:2005ze} 
%that incoherent production leads to excitation
%of the target nucleus which decays by emission of neutrons.
%
Despite large experimental errors, it seems that the agreement between our calculations and the 
PHENIX data demonstrated in Fig.~\ref{phenix} justifies our approach to neutron production in UPCs.
Coupling this with the good description of coherent $J/\psi$ photoproduction at the LHC makes our 
approach reasonable for the prediction of  coherent and incoherent UPCs 
accompanied by neutron emission at the LHC.

\begin{figure}[t]
\centering
\epsfig{file=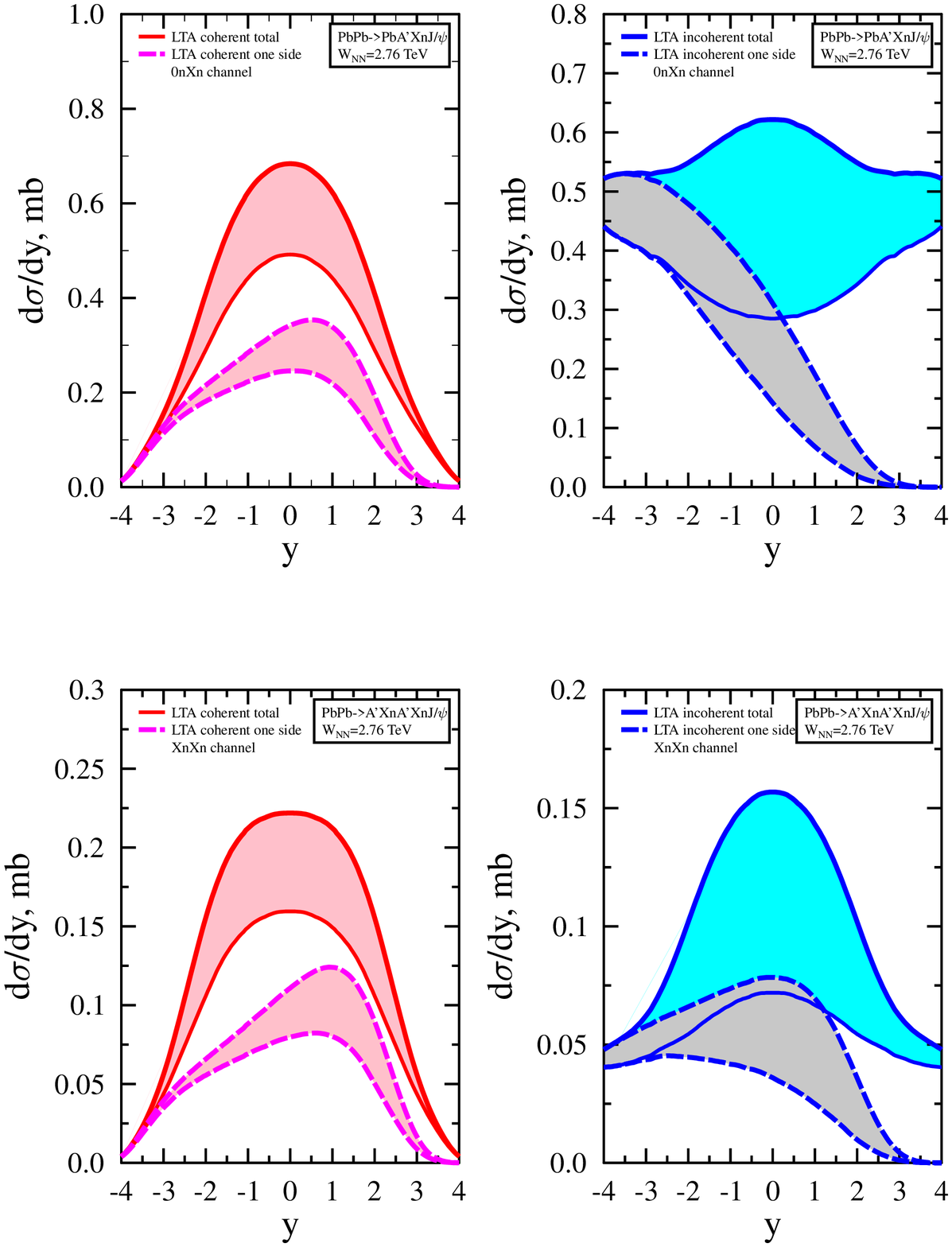,scale=0.65}
\caption{Predictions of the leading twist approximation (LTA) for the rapidity distribution of 
coherent (two left panels) and incoherent (two right panels) $J/\psi$ photoproduction in Pb-Pb UPCs with neutron emission in the 
0nXn-channel (two upper panels) and in the XnXn-channel (two lower panels) at $\sqrt{s}=2.76$ GeV.
%vg
The shaded bands represent the theoretical uncertainty of the LTA predictions.
}
\label{cms}
\end{figure}

Figure~\ref{cms}
presents the results of our calculation of the rapidity distributions 
in $J/\psi$ photoproduction in Pb-Pb UPCs  with neutron emission in the 
0nXn-channel (upper row of panels) and the XnXn-channel (lower row of panels) as functions of the 
rapidity $y$
at the LHC energy of ${\sqrt s_{NN}}=2.76$ GeV;
the two panels on the left correspond to the coherent case and the panels on the right 
are for the incoherent case.
Our theoretical predictions are made using the leading twist approximation (LTA); 
the shaded areas represent the dominant LTA theoretical uncertainty 
%vg2014
associated with the uncertainty in the predicted nuclear gluon distributions~\cite{Frankfurt:2011cs}.
The uncertainty associated with the value of the hard scale $\mu^2$ is much smaller and, hence, has been 
neglected.

Each panel contains two sets of curves: 
the upper shaded area is a sum of the two terms 
in Eq.~(\ref{csupc}) and the lower shaded area represents the contribution of $J/\psi$ production 
by the photon emitted by the nucleus moving with the momentum in the direction of positive 
$J/\psi$  rapidity (the dashed curves labeled ``one side'').

We can draw several conclusions from Fig.~\ref{cms}.
First, one can see from 
the two left panels
that using the data 
on the 0nXn and XnXn-channels,
one can try to extract the high-photon-energy contribution using Eq.~(\ref{eq:disentangling}).
In the range of rapidities of $1.5<y<2.5$, this will allow one 
to determine nuclear gluon shadowing in Pb down to $x\approx 10^{-4}$ and at the scale of
$\mu ^2 \approx 3$ GeV$^2$
from coherent $J/\psi$ photoproduction
in UPCs. 

Second, a comparison of the corresponding upper shaded areas  
in Fig.~\ref{cms}
shows that we predict that at central rapidities, 
(i) in the 0nXn-channel, the coherent and incoherent contributions will be
practically of the same magnitude and (ii) in the  XnXn-channel, 
the coherent contribution exceeds the incoherent one by approximately a factor of two.

Third, besides the coherent channel, the incoherent cross section can also be used to 
study nuclear gluon shadowing at small $x$. Based on the dominance of the
strong interaction mechanism of neutron production in incoherent photoproduction,
we predict that there is a good opportunity to separate the low-photon-energy and 
high-photon-energy contributions  in the 0nXn-channel 
(see the upper right panel in Fig.~\ref{cms}).
Indeed, since in this case neutrons are 
emitted by the target nucleus, there should by a correlation between the direction
of the produced $J/\psi$ and the direction of neutrons in UPC events. 
In the kinematics of UPCs, the 
direction of charmonium
produced by a high-energy photon and, hence,  low-$x$ gluons from the target, is opposite
to the direction of the target nucleus and, correspondingly, to the 
direction of neutrons. 
Conversely, in the low-photon-energy production, the direction
of charmonium coincides with the direction of the target and neutrons.

\begin{figure}[t]
\centering
\epsfig{file=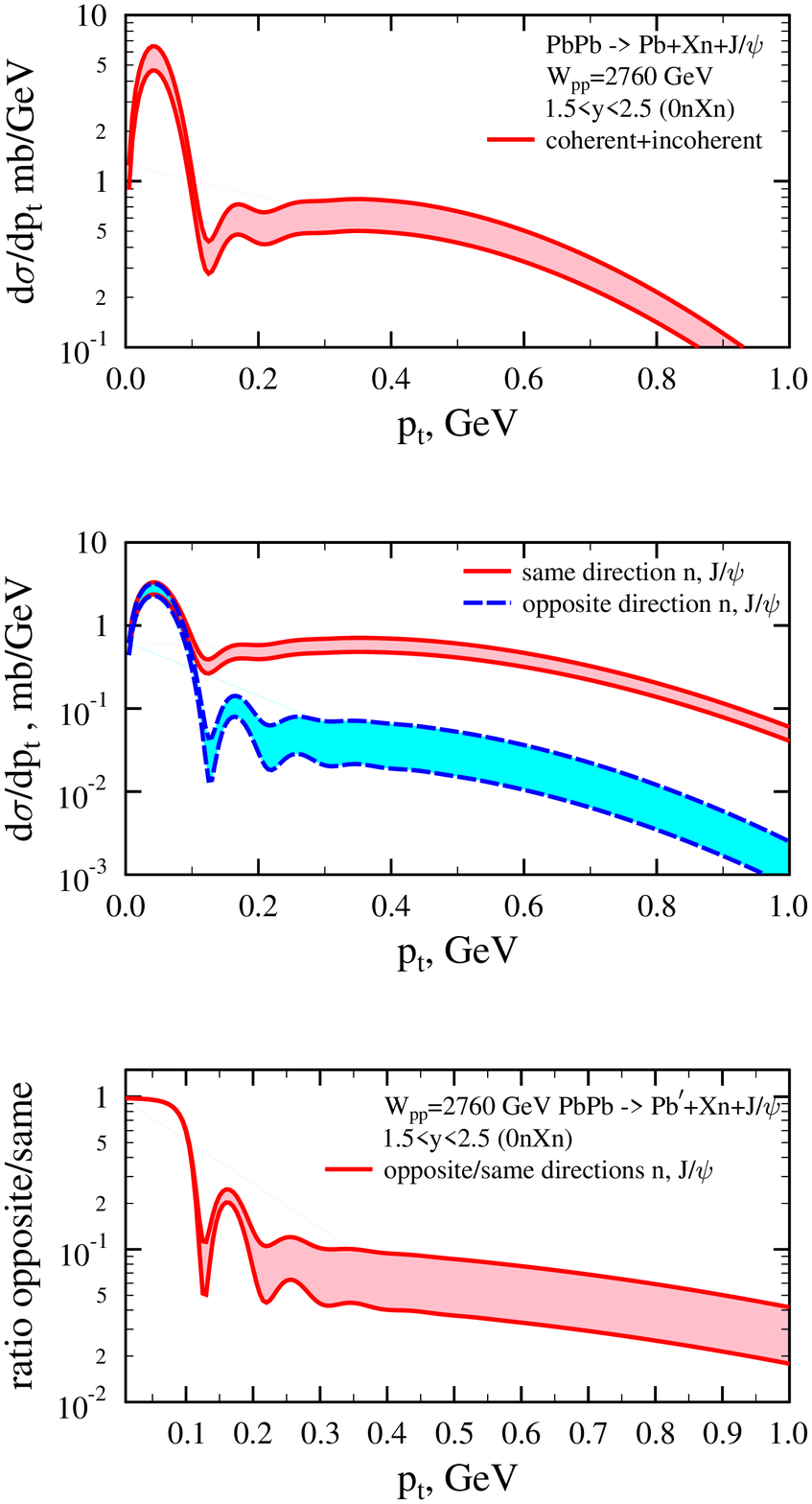,scale=0.6}
\caption{Upper panel: Predictions for the sum of the coherent and incoherent cross section of $J/\psi$ photoproduction in Pb-Pb UPCs accompanied by neutron emission in the 0nXn-channel 
as a function of $p_t$ in the rapidity range of  $1.5<y<2.5$.
Middle panel: The same quantity for the situation when the direction of $J/\psi$ coincides with that
of the neutrons (upper band) and when the directions of $J/\psi$
are the neutrons are opposite (lower band). Bottom panel: the ratio of the two curves from the middle panel. 
The shaded bands represent the theoretical uncertainty of the LTA predictions.
}
\label{ptdist}
\end{figure}

The standard procedure of the separation of coherent and incoherent
events consists in the analysis of momentum transfer distributions. In coherent photoproduction,
this distribution is dictated by the nuclear form factor squared and presents 
several distinct peaks
at small $p_t <200$ MeV/c, where $p_t$ is the transverse momentum 
of produced $J/\psi$ and $t=-p_t^2$.
One can see such first three diffractive peaks in  Fig.~\ref{ptdist}. 
 The $t$ dependence of the incoherent nuclear cross section 
is the same as in the elementary process, i.e., 
$\exp [-B_{J/\psi} |t|]$. 
Our predictions for the sum of the coherent and incoherent cross sections of $J/\psi$ photoproduction 
in Pb-Pb UPCs accompanied by neutron emission in the 0nXn-channel 
as a function of $p_t$ in the rapidity range of  $1.5<y<2.5$
 is shown in the upper panel of Fig.~\ref{ptdist}. As in Fig.~\ref{cms},
our 
theoretical predictions are based on the LTA;
%vg2014
the shaded bands represent the LTA uncertainty in the predicted nuclear gluon distribution.
In the figure, the peaks at small $p_t$ correspond to the coherent signal. Therefore, 
the $p_t <200$ MeV/c cut will effectively reject incoherent events with good accuracy.

As we already discussed above,
the incoherent cross section can also probe the small-$x$ nuclear gluon distribution.
In particular, in the 0nXn-channel, the directions of $J/\psi$ and the emitted neutrons can be 
unambiguously correlated with either the low-photon-energy or with the high-photon-energy
contributions.
At the same time, in the coherent channel,
due to the assumption of independence of the processes of
coherent photoproduction and electromagnetic dissociation followed by the nucleus neutron decay,
it is impossible to assign emitting neutrons to the source or the target, and,
hence, the probabilities to detect neutrons in the same and in the opposite
direction with respect to the direction of $J/\psi$ 
should be equal. 

These predictions are demonstrated in the middle panel of Fig.~\ref{ptdist}, 
where the upper band corresponds to the situation when the direction of $J/\psi$ coincides with that
of the neutrons and the lower band is for the opposite directions of $J/\psi$ and the neutrons.
The bottom panel shows the ratio of the two curves from the middle panel. 
Since we identify the events, where the directions of $J/\psi$ and neutrons are opposite,
with high-energy photoproduction, there could be two sources of suppression of the corresponding 
cross section:
 (i) the falloff of the flux of high-energy photons emitted by the nucleus 
and (ii) stronger nuclear gluon shadowing of the small-$x$ gluons at 
$x \approx 10^{-4}$,
which we are mainly interested in.

\section{Conclusions}
\label{sec:conclusions}

We considered $J/\psi$ photoproduction in ion--ion UPCs at the LHC and RHIC in the coherent
and incoherent channels with and without accompanying forward neutron emission and analyzed the role
of nuclear gluon shadowing at small $x$, $x=10^{-4}-10^{-2}$, in these processes.
We extended the formalism of leading twist nuclear shadowing characterized by large nuclear
gluon shadowing to the incoherent $\sigma_{\gamma A \to J/\psi A^{\prime}}$ cross section. We found 
that despite a good agreement between the approaches predicting large nuclear gluon shadowing 
at 
%vg $x=10^{-3}-10^{-2}$
$x \approx 10^{-3}$
and the large nuclear suppression factor extracted from the ALICE data on coherent $J/\psi$ 
photoproduction in Pb-Pb UPCs at the LHC~\cite{Abbas:2013oua,Abelev:2012ba},
in the incoherent channel,
  the leading twist 
approximation predicts the amount of nuclear suppression due to gluon shadowing which 
exceeds that seen in the data by approximately a factor of $1.5$ (Fig.~\ref{cirapalice}). 
We hypothesize that one source of the discrepancy 
could be the contribution of incoherent nucleon dissociation, $\gamma N \to J/\psi Y$, which 
%vgis implicitly included in 
could potentially contribute to
the ALICE data~\cite{Abbas:2013oua} and which is not taken into account in our 
theoretical analysis.

In coherent $J/\psi$ photoproduction in ion--ion UPCs, it is problematic (except for $y \approx 0$ and
large $|y|$) to separate the contributions of high-energy and low-energy photons to the 
$d \sigma_{AA\to AAJ/\psi}(y)/dy$ cross section, which reduces the range of $x$ for the studies of
small-$x$ nuclear gluon shadowing. This problem can be circumvented by considering  
 $J/\psi$ photoproduction in ion--ion UPCs accompanied by neutron emission due to electromagnetic
excitation of one or both colliding nuclei.

Using the leading twist approximation for nuclear gluon shadowing, 
we made predictions for coherent and incoherent nuclear $J/\psi$ photoproduction in Pb-Pb UPCs 
accompanied by neutron emission in various channels at the LHC (Fig.~\ref{cms}).
In particular, we discuss the strategy allowing one to separate the low-photon-energy and
the high-photon-energy contributions to coherent $J/\psi$ photoproduction performing
a joint analysis of the data in the 0nXn and XnXn-channels. 
This gives an opportunity to shift the study of nuclear gluon shadowing to the lower $x$ region 
of  $x \approx 10^{-4}$.

In addition, in the incoherent case accompanied by neutron emission, 
we show that the separation between the low-photon-energy and
high-photon-energy contributions can be efficiently performed by measuring the correlation between 
the directions of $J/\psi$ and the emitted neutrons (Fig.~\ref{ptdist}).

In the kinematics where nuclear shadowing is small,
we showed (Fig.~\ref{phenix}) that theoretical predictions based on the dipole model agree with the PHENIX data on 
$J/\psi$ photoproduction in Au-Au UPCs at $\sqrt{s_{NN}}=200$ GeV in the XnXn-channel 
(both colliding nuclei emit neutrons detected in the ZDCs).

\section*{Acknowledgments}
\label{sec:acknowledgments}

We would like to thank L.~Frankfurt for useful discussions. 
M.~Strikman's research was supported by DOE grant No.~DE-FG02-93ER40771.


\begin{thebibliography}{99}


%\cite{Abbas:2013oua}
\bibitem{Abbas:2013oua} 
  E.~Abbas {\it et al.}  [ALICE Collaboration],
  %``Charmonium and $e^+e^-$ pair photoproduction at mid-rapidity in ultra-peripheral Pb-Pb collisions at $\sqrt{s_{NN}}$ = 2.76 TeV,''
  Eur.\ Phys.\ J.\ C {\bf 73}, 2617 (2013)
  [arXiv:1305.1467 [nucl-ex]].
  %%CITATION = ARXIV:1305.1467;%%
  %13 citations counted in INSPIRE as of 20 Dec 2013

\bibitem{Abelev:2012ba} 
  B.~Abelev {\it et al.}  [ALICE Collaboration],
  %``Coherent $J/\psi$ photoproduction in ultra-peripheral Pb-Pb collisions at $\sqrt{s_{NN}} = 2.76$ TeV,''
  Phys.\ Lett.\ B {\bf 718}, 1273 (2013)
  [arXiv:1209.3715 [nucl-ex]]. 


\bibitem{Guzey:2013xba} 
  V.~Guzey, E.~Kryshen, M.~Strikman and M.~Zhalov,
  %``Evidence for nuclear gluon shadowing from the ALICE measurements of PbPb ultraperipheral exclusive $J/{\psi}$ production,''
  Phys.\ Lett.\ B {\bf 726}, 290 (2013)
  [arXiv:1305.1724 [hep-ph]]. 
  
%\bibitem{Guzey:2013qza} 
%  V.~Guzey and M.~Zhalov,
  %``Exclusive J/{\psi} production in ultraperipheral collisions at the LHC: constrains on the gluon distributions in the proton and nuclei,''
%  arXiv:1307.4526 [hep-ph].
%\cite{Guzey:2013qza}
\bibitem{Guzey:2013qza} 
  V.~Guzey and M.~Zhalov,
  %``Exclusive $J/{\psi}$ production in ultraperipheral collisions at the LHC: constrains on the gluon distributions in the proton and nuclei,''
  JHEP {\bf 1310}, 207 (2013)
  [arXiv:1307.4526 [hep-ph]].
  %%CITATION = ARXIV:1307.4526;%%
  %2 citations counted in INSPIRE as of 24 Nov 2013

%\cite{Eskola:2009uj}
\bibitem{Eskola:2009uj} 
  K.~J.~Eskola, H.~Paukkunen and C.~A.~Salgado,
  %``EPS09: A New Generation of NLO and LO Nuclear Parton Distribution Functions,''
  JHEP {\bf 0904}, 065 (2009)
  [arXiv:0902.4154 [hep-ph]].
  %%CITATION = ARXIV:0902.4154;%%
  %293 citations counted in INSPIRE as of 24 Nov 2013

\bibitem{Frankfurt:2011cs}
  L.~Frankfurt, V.~Guzey and M.~Strikman,
  %``Leading Twist Nuclear Shadowing Phenomena in Hard Processes with Nuclei,''
  Phys.\ Rept.\  {\bf 512}, 255 (2012)
  [arXiv:1106.2091 [hep-ph]].
  %%CITATION = ARXIV:1106.2091;%%

\bibitem{Frankfurt:2008et} 
  L.~Frankfurt, M.~Strikman and M.~Zhalov,
  %``Tracking fast small color dipoles through strong gluon fields at the LHC,''
  Phys.\ Rev.\ Lett.\  {\bf 102}, 232001 (2009)
  [arXiv:0811.0368 [hep-ph]].
  %%CITATION = ARXIV:0811.0368;%%
  %6 citations counted in INSPIRE as of 12 Mar 2014

%\cite{Sakurai_1969}
\bibitem{Sakurai_1969} 
  J.~J.~Sakurai,
  %``Vector meson dominance and high-energy electron proton inelastic scattering,''
  Phys.\ Rev.\ Lett.\  {\bf 22}, 981 (1969).
  %%CITATION = PRLTA,22,981;%%
  %244 citations counted in INSPIRE as of 03 Dec 2013

\bibitem{Bauer:1977iq} 
  T.~H.~Bauer, R.~D.~Spital, D.~R.~Yennie and F.~M.~Pipkin,
  %``The Hadronic Properties of the Photon in High-Energy Interactions,''
  Rev.\ Mod.\ Phys.\  {\bf 50}, 261 (1978)
  [Erratum-ibid.\  {\bf 51}, 407 (1979)].

%\cite{Glauber:1955qq}
\bibitem{Glauber:1955qq}
  R.~J.~Glauber,
  %``Cross-sections in deuterium at high-energies,''
  Phys.\ Rev.\  {\bf 100}, 242 (1955).
  %%CITATION = PHRVA,100,242;%%

%\cite{Hufner:1997jg}
\bibitem{Hufner:1997jg} 
  J.~Hufner and B.~Z.~Kopeliovich,
  %``J / Psi N and Psi-prime N total cross-sections from photoproduction data: Failure of vector dominance,''
  Phys.\ Lett.\ B {\bf 426}, 154 (1998)
  [hep-ph/9712297].
  %%CITATION = HEP-PH/9712297;%%
  %47 citations counted in INSPIRE as of 24 Nov 2013


%\cite{Nikolaev:1990ja}
\bibitem{Nikolaev:1990ja} 
  N.~N.~Nikolaev and B.~G.~Zakharov,
  %``Color transparency and scaling properties of nuclear shadowing in deep inelastic scattering,''
  Z.\ Phys.\ C {\bf 49}, 607 (1991).
  %%CITATION = ZEPYA,C49,607;%%
  %727 citations counted in INSPIRE as of 04 Dec 2013

%\cite{Mueller:1994jq}
\bibitem{Mueller:1994jq} 
  A.~H.~Mueller and B.~Patel,
  %``Single and double BFKL pomeron exchange and a dipole picture of high-energy hard processes,''
  Nucl.\ Phys.\ B {\bf 425}, 471 (1994)
  [hep-ph/9403256].
  %%CITATION = HEP-PH/9403256;%%
  %506 citations counted in INSPIRE as of 20 Dec 2013

%\cite{Blaettel:1993rd}
\bibitem{Blaettel:1993rd} 
  B.~Blaettel, G.~Baym, L.~L.~Frankfurt and M.~Strikman,
  %``How transparent are hadrons to pions?,''
  Phys.\ Rev.\ Lett.\  {\bf 70}, 896 (1993).
  %%CITATION = PRLTA,70,896;%%
  %89 citations counted in INSPIRE as of 20 Dec 2013

%
\bibitem{Ryskin:1992ui}
  M.~G.~Ryskin,
  %``Diffractive J / psi electroproduction in LLA QCD,''
  Z.\ Phys.\  {\bf C57}, 89 (1993).

\bibitem{Brodsky:1994kf}
  S.~J.~Brodsky, L.~Frankfurt, J.~F.~Gunion, A.~H.~Mueller and M.~Strikman,
  %``Diffractive leptoproduction of vector mesons in QCD,''
  Phys.\ Rev.\  D {\bf 50}, 3134 (1994). 

%\cite{Frankfurt:2013ria}
%\bibitem{Frankfurt:2013ria}
%  L.~Frankfurt and M.~Strikman,
  %``Diffractive phenomena in high energy processes,''
%  arXiv:1304.4308 [hep-ph].
  %%CITATION = ARXIV:1304.4308;%%

%\cite{Kopeliovich:2001xj}
\bibitem{Kopeliovich:2001xj} 
  B.~Z.~Kopeliovich, J.~Nemchik, A.~Schafer and A.~V.~Tarasov,
  %``Color transparency versus quantum coherence in electroproduction of vector mesons off nuclei,''
  Phys.\ Rev.\ C {\bf 65}, 035201 (2002)
  [hep-ph/0107227].
  %%CITATION = HEP-PH/0107227;%%
  %75 citations counted in INSPIRE as of 18 Dec 2013


%\cite{Goncalves:2011vf}
\bibitem{Goncalves:2011vf}
  V.~P.~Goncalves and M.~V.~T.~Machado,
%  {\it Vector Meson Production in Coherent Hadronic Interactions: An update on predictions for RHIC and LHC},
  Phys.\ Rev.\ C {\bf 84}, 011902 (2011) 
  [arXiv:1106.3036 [hep-ph]].
  %%CITATION = ARXIV:1106.3036;%%
  %17 citations counted in INSPIRE as of 09 Jul 2013

%\cite{Lappi:2013am}
\bibitem{Lappi:2013am}
  T.~Lappi and H.~Mantysaari,
%  {\it $J/Psi$ production in ultraperipheral Pb+Pb and p+Pb collisions at LHC energies},
  Phys.\ Rev.\ C {\bf 87}, 032201 (2013)
 [arXiv:1301.4095 [hep-ph]];
  %%CITATION = ARXIV:1301.4095;%%
  %5 citations counted in INSPIRE as of 16 Jul 2013
%\cite{Lappi:2013jgs}
%\bibitem{Lappi:2013jgs} 
%  T.~Lappi and H.~Mäntysaari,
  %``Models for exclusive vector meson production in heavy-ion collisions,''
  arXiv:1310.6336 [hep-ph].
  %%CITATION = ARXIV:1310.6336;%%



%\cite{Frankfurt:2002kd}
\bibitem{Frankfurt:2002kd} 
  L.~Frankfurt, V.~Guzey, M.~McDermott and M.~Strikman,
  %``Nuclear shadowing in deep inelastic scattering on nuclei: Leading twist versus eikonal approaches,''
  JHEP {\bf 0202}, 027 (2002)
  [hep-ph/0201230].
  %%CITATION = HEP-PH/0201230;%%
  %51 citations counted in INSPIRE as of 04 Dec 2013
 

%\cite{Gribov_1969}
\bibitem{Gribov_1969} 
  V.~N.~Gribov,
  %``Glauber corrections and the interaction between high-energy hadrons and nuclei,''
  Sov.\ Phys.\ JETP {\bf 29}, 483 (1969)
  [Zh.\ Eksp.\ Teor.\ Fiz.\  {\bf 56}, 892 (1969)].
  %%CITATION = SPHJA,29,483;%%
  %419 citations counted in INSPIRE as of 05 Dec 2013

%\cite{Abramovsky:1973fm}
\bibitem{Abramovsky:1973fm}
  V.~A.~Abramovsky, V.~N.~Gribov and O.~V.~Kancheli,
  %``CHARACTER OF INCLUSIVE SPECTRA AND FLUCTUATIONS PRODUCED IN INELASTIC
  %PROCESSES BY MULTI - POMERON EXCHANGE,''
  Yad.\ Fiz.\  {\bf 18}, 595 (1973)
  [Sov.\ J.\ Nucl.\ Phys.\  {\bf 18}, 308 (1974)];
  %%CITATION = SJNCA,18,308;%%
also in V.N. Gribov, {\it Gauge theories and quark confinement}, Phasis, Moskow, 2002, 
p. 67.

  
%\cite{Good_1960}
\bibitem{Good_1960} 
  M.~L.~Good and W.~D.~Walker,
  %``Diffraction disssociation of beam particles,''
  Phys.\ Rev.\  {\bf 120}, 1857 (1960).
  %%CITATION = PHRVA,120,1857;%%
  %383 citations counted in INSPIRE as of 05 Dec 2013


%\cite{Collins:1997sr}
\bibitem{Collins:1997sr}
  J.~C.~Collins,
  %``Proof of factorization for diffractive hard scattering,''
  Phys.\ Rev.\  D {\bf 57}, 3051 (1998)
  [Erratum-ibid.\  D {\bf 61}, 019902 (2000)].
%  [arXiv:hep-ph/9709499].
  %%CITATION = PHRVA,D57,3051;%%

%\cite{Aktas:2006hy}
\bibitem{Aktas:2006hy}
  A.~Aktas {\it et al.}  [H1 Collaboration],
  %``Measurement and QCD analysis of the diffractive deep-inelastic  scattering
  %cross section at HERA,''
  Eur.\ Phys.\ J.\  C {\bf 48}, 715 and 749 (2006);\\
%  [arXiv:hep-ex/0606004].
  %%CITATION = EPHJA,C48,715;%%
%\cite{Aktas:2006hx}
%\bibitem{Aktas:2006hx}
%  A.~Aktas {\it et al.}  [H1 Collaboration],
  %``Diffractive deep-inelastic scattering with a leading proton at HERA,''
%  Eur.\ Phys.\ J.\  C {\bf 48}, 749 (2006).
%  [arXiv:hep-ex/0606003].
  %%CITATION = EPHJA,C48,749;%%
%\cite{Chekanov:2009qja}
%\bibitem{Chekanov:2009qja}
  S.~Chekanov {\it et al.}  [ZEUS Collaboration],
  %``A QCD analysis of ZEUS diffractive data,''
  Nucl.\ Phys.\  B {\bf 831}, 1 (2010).
%  [arXiv:0911.4119 [hep-ex]].
  %%CITATION = NUPHA,B831,1;%%

%\cite{Afanasiev:2009hy}
\bibitem{Afanasiev:2009hy}
  S.~Afanasiev {\it et al.}  [PHENIX Collaboration],
  %``Photoproduction of J/psi and of high mass e+e- in ultra-peripheral Au+Au collisions at s**(1/2) = 200-GeV,''
  Phys.\ Lett.\ B {\bf 679}, 321 (2009) 
  [arXiv:0903.2041 [nucl-ex]].
  %%CITATION = ARXIV:0903.2041;%%
  %32 citations counted in INSPIRE as of 07 Mar 2014
  

%\cite{Motyka:2008jk}
\bibitem{Motyka:2008jk} 
  L.~Motyka, K.~Golec-Biernat and G.~Watt,
  %``Dipole models and parton saturation in ep scattering,''
  arXiv:0809.4191 [hep-ph].
  %%CITATION = ARXIV:0809.4191;%%
  %18 citations counted in INSPIRE as of 07 Mar 2014


%\cite{Rezaeian:2012ji}
\bibitem{Rezaeian:2012ji} 
  A.~H.~Rezaeian, M.~Siddikov, M.~Van de Klundert and R.~Venugopalan,
  %``Analysis of combined HERA data in the Impact-Parameter dependent Saturation model,''
  Phys.\ Rev.\ D {\bf 87}, no. 3, 034002 (2013)
  [arXiv:1212.2974].
  %%CITATION = ARXIV:1212.2974;%%
  %16 citations counted in INSPIRE as of 07 Mar 2014

\bibitem{Baltz:2007kq} 
  A.~J.~Baltz {\it et al.},
 %G.~Baur, D.~d'Enterria, L.~Frankfurt, F.~Gelis, V.~Guzey, K.~Hencken and Y.~.Kharlov {\it et al.},
  %``The Physics of Ultraperipheral Collisions at the LHC,''
  Phys.\ Rept.\  {\bf 458}, 1 (2008)
  [arXiv:0706.3356 [nucl-ex]].


%\cite{Pumplin:2002vw}
\bibitem{Pumplin:2002vw}
  J.~Pumplin, D.~R.~Stump, J.~Huston, H.~L.~Lai, P.~M.~Nadolsky and W.~K.~Tung,
%  {\it New generation of parton distributions with uncertainties from global QCD analysis},
  JHEP {\bf 0207}, 012 (2002) 
  [hep-ph/0201195].
  %%CITATION = HEP-PH/0201195;%%
  %3346 citations counted in INSPIRE as of 12 May 2013

\bibitem{Frankfurt:2008er} 
  L.~Frankfurt, M.~Strikman and M.~Zhalov,
  %``Large t diffractive J/psi photoproduction with proton dissociation in ultraperipheral pA collisions at LHC,''
  Phys.\ Lett.\ B {\bf 670}, 32 (2008)
  [arXiv:0807.2208 [hep-ph]].

%\cite{Alexa:2013xxa}
\bibitem{Alexa:2013xxa} 
  C.~Alexa {\it et al.}  [H1 Collaboration],
  %``Elastic and Proton-Dissociative Photoproduction of J/psi Mesons at HERA,''
  Eur.\ Phys.\ J.\ C {\bf 73}, 2466 (2013)
  [arXiv:1304.5162 [hep-ex]].
  %%CITATION = ARXIV:1304.5162;%%
  %8 citations counted in INSPIRE as of 18 Dec 2013


  \bibitem{Rebyakova:2011vf} 
  V.~Rebyakova, M.~Strikman and M.~Zhalov,
  %``Coherent rho and J/psi photoproduction in ultraperipheral processes with electromagnetic dissociation of heavy ions at RHIC and LHC,''
  Phys.\ Lett.\ B {\bf 710}, 647 (2012)
  [arXiv:1109.0737 [hep-ph]].
 
\bibitem{Baltz:2002pp} 
  A.~J.~Baltz, S.~R.~Klein and J.~Nystrand,
  %``Coherent vector meson photoproduction with nuclear breakup in relativistic heavy ion collisions,''
  Phys.\ Rev.\ Lett.\  {\bf 89}, 012301 (2002)
  [nucl-th/0205031].


\bibitem{pshenichnov}
I.~Pshenichnov {\it et. al.}, 
Phys.\ Rev.\ C {\bf 64}, 024903 (2001); \\
I.~Pshenichnov,
Phys.\ Part.\ Nucl.\ {\bf 42}, 215 (2011).

\bibitem{ALICE:2012aa} 
  B.~Abelev {\it et al.}  [ALICE Collaboration],
  %``Measurement of the Cross Section for Electromagnetic Dissociation with Neutron Emission in Pb-Pb Collisions at $\sqrt{s_{NN}}$ = 2.76 TeV,''
  Phys.\ Rev.\ Lett.\  {\bf 109}, 252302 (2012)
  [arXiv:1203.2436 [nucl-ex]].


\bibitem{Strikman:2005ze} 
  M.~Strikman, M.~Tverskoy and M.~Zhalov,
  %``Neutron tagging of quasielastic J/psi photoproduction off nucleus in ultraperipheral heavy ion collisions at RHIC energies,''
  Phys.\ Lett.\ B {\bf 626}, 72 (2005)
  [hep-ph/0505023].


\bibitem{takahara}
  A.~Takahara for the PHENIX collaboration,
  % "$J/\psi$ photoproduction in ultraperipheral Au+Au collisions at
  % $\sqrt {s_{NN}}=200$ GeV measured by RHIC-PHENIX collaboration"
  Proceedings of XX International Workshop on Deep Inelastic Scattering 2012.
  University of Bonn, 26-30 June, 2012,
  DOI: http://dx.doi.org/10.3204/DESY-PROC-2012-02/114.
  %Proceedings of International Workshop on High Energy Scattering
  %at Zero Degree, Nagoya, March, 2013.  
  %http://www.gcoe.phys.nagoya-u.ac.jp/hesz2013/program.html
 
  
  \bibitem{GolecBiernat:1998js} 
  K.~J.~Golec-Biernat and M.~Wusthoff,
  %``Saturation effects in deep inelastic scattering at low Q**2 and its implications on diffraction,''
  Phys.\ Rev.\ D {\bf 59}, 014017 (1998).
  
\end{thebibliography}
\end{document}